\documentclass{sig-alternate}
\usepackage{array}          
\usepackage{amsmath}        
\usepackage{amssymb}        
\usepackage{mathrsfs}       
\usepackage{wrapfig}
\usepackage{url}
\usepackage{fixltx2e}
\usepackage{breakcites}
\usepackage{placeins}
\usepackage{color}
\usepackage{colortbl}
\usepackage[breaklinks]{hyperref}
\hypersetup{
    colorlinks=true,       
    linkcolor=blue,          
    citecolor=black,        
    filecolor=black,      
    urlcolor=blue           
}
\usepackage{multirow}    
\usepackage{caption}
\usepackage{subcaption}    
\begin{document}
%
%
\title{Mobile Homophily and Social Location Prediction}
%
\numberofauthors{3}
\author{
\alignauthor
Halgurt Bapierre\\
       \affaddr{TU M\"unchen}\\
       \affaddr{Faculty for Informatics}\\
       \email{halgurt@gmx.de}
\alignauthor
Chakajkla Jesdabodi\\
       \affaddr{TU M\"unchen}\\
       \affaddr{Faculty for Informatics}\\
       \email{c.jesdabodi@gmail.com}
\alignauthor
Georg Groh\\
       \affaddr{TU M\"unchen}\\
       \affaddr{Faculty for Informatics}\\
       \email{grohg@in.tum.de}
}
\maketitle
\begin{abstract}
The mobility behavior of human beings is predictable to a varying degree e.g. depending on the traits of their personality such as the trait extraversion - introversion: 
the mobility of introvert users may be more dominated by routines and habitual movement patterns, resulting in a more predictable mobility behavior on the basis of their own location history while, in contrast, extrovert users get about a lot and are explorative by nature, which may hamper the prediction of their mobility. However, socially more active and extrovert users meet more people and share information, experiences, believes, thoughts etc. with others. which in turn leads to a high interdependency between their mobility and social lives. Using a large LBSN dataset, his paper investigates the interdependency between human mobility and social proximity, the influence of social networks on enhancing location prediction of an individual and the transmission of social trends/influences within social networks.
\end{abstract}
%
\section{Introduction}
\label{intro}
General (social) homophily refers to the tendency of humans to socially connect to other individuals with similar personal properties \cite{McPherson2001}. If similarity is evaluated with respect to geographical distance (e.g. of the center of living), this tendency may be referred to as Propinquity \cite{Kadushin2012}. If the similarity is evaluated with respect to locations visited and/or the temporal sequence of these visits, we speak of mobile homophily. More technically, for the rest of this contribution, mobile homophily will refer to similarity between users with respect to their mobility behavior. We investigate the relation between human social relations and mobile homophily with the ultimate goal of improving next location predictions for users using social network data. In a first part we focus on studying the correlations between social relations and mobile homophily, using a large dataset from a location based social network (LBSN). Here, we especially focus studying the effects of tie strength and dense sub-groups. In a second part we study in how far we can exploit these correlations for improving next location predictions on the basis of data on user's social relations.
\section{Social Relations and Mobile Homophily: Related Work}\label{homophily}
Social relations and geographic distance $d$ exhibit many interesting interrelations. Propinquity has been studied in form of the probability of friendship relations as a function of $d$: \cite{LucillePowell1975, Fischer1997, liben2005geographic, lambiotte2008, backstrom2010, scellato2011, scellato2011c, Volkovich2012}. Most studies find a power law relation $p(d)\propto d^{-\alpha}$ with slightly different exponents.
\cite{Volkovich2012} find an inverse correlation between distance of centers of life of two users and the relative size of the overlap of their immediate social relations.
In contrast to that, \cite{Kaltenbrunner2012} found that purely online (virtual) interaction between users may not be not strongly influenced by distance.
The mutual influence of mobility and social tie strength has also been investigated by \cite{barabasi2011, eunjoon2011}
\cite{scellato2011c} show that a substantial share of new friendships may be predicted from co-location events of users.
\cite{gonzales2006} were able to reconstruct a social network via an analysis of the motion patterns of the corresponding users.
In view of social link prediction \cite{liben2007}, \cite{Scellato2011b} show that $\approx$30\% of new friendships are formed between persons that have visited at least one common location.
Regarding mobile homophily, \cite{Dong2011, Crandall2010} show that the probability of two users having social ties increases with more number of co-locations and reciprocally there is an increase in shared location of friends as time grows.
\cite{Cranshaw2010, Noulas2011b} investigate further relations between mobility behavior and social relationships, taking the category of the locations into account. \cite{Cranshaw2010} also regard the regularity of the mobility patterns.
Using a large mobile phone based data-set with cell-tower-based localization granularity, \cite{barabasi2011} also investigate  social ties and social tie strength in relation to the mobility patterns of the respective users. We will refer to elements of this study later in more depth.

However, most of the studies do not explicitly focus on the influence of social relationships on next location prediction. Before we focus on the investigation of these influences in \autoref{socialPrediction} we will discuss the results of our study on the relation between social network and mobility behavior using our Foursquare data-set which we will now discuss.
\section{Data-Set} \label{dataset}
In contrast to studies using cell-tower- or Bluetooth-based location and co-location inference \cite{MadanPentland2009, barabasi2011}, Location Based Social Networking (LBSN) platforms provide GPS-accurate and explicit declarations of visits of users to locations (``check-ins'') and more detailed data on the nature of these locations. Compare \cite{eunjoon2011,Brown2012,Gao2012} for recent studies using LBSN data-sets. LBSN check-ins do not allow to directly detect social location visits but we may heuristically infer these co-location visits with an accuracy comparable to Bluetooth encounters via their time-stamps. The time stamps of check-ins also allow to investigate e.g. the dynamics of influence of users on other users \cite{Goyal2010,Sadilek2012} with respect to location behavior \cite{weiPan2012,Wang2011b,Altshuler2011} and allow for assessments of the social influence on next location prediction.

As an LBSN to collect a data-set we used Foursquare \cite{FoursquareAbout}, providing fine grained locations with time-stamped check-ins and social networking services for users, allowing to construct a social network. We restricted the collection to all check-ins of all venues of the San-Francisco area because Foursquare has a large active user-base in that area and thus we assumed that the collected check ins represent a temporally, spatially and socially sufficiently dense cover of the actual mobility of the involved users.

For the social network, we extract the friends and the friends of friends of all users who have made at least one check-in within the four month period (122 days) of data collection. We refer to the set of users how have generated at least 50 check-ins within the course of data collection period with the set of active users. We use the set of active users for investigating human mobility behavior, because of the availability of sufficient location data. \autoref{table1} contains descriptive statistics for the data-set. The average degree is higher and the mean average path length is shorter for the social networks containing only the active users, their friends and their friends of friends, because the active users are on Foursquare since a longer period of time, otherwise they would not have generated more check-ins on average compared to all users (\autoref{table1}).

The Foursquare social network is a hybrid social network containing both real world and online friends, thus, the average degree of (96) in our dataset is lower compared to pure online social networks such as Facebook with an average degree of (190) \cite{AnatomyFacebook}. Further, the mean average path lengths in both social networks with all users and active users are found to be 4.152 for our dataset and 3.8 for Facebook \cite{AnatomyFacebook} respectively, which is also in line with Stanley Milgram's small world phenomenon \cite{Milgram1967, Travers1969}.

\begin{table}[htb]
\begin{center}
\begin{tabular}[c]{|l|l|l|}
\hline
\footnotesize{\textbf{Quantity}} & \footnotesize{\textbf{all}} & \footnotesize{\textbf{active}}\\
\hline \hline
  \scriptsize{\# nodes (users): $U$ } & \scriptsize 141,750 & \scriptsize 9173  \\
  \hline
  \scriptsize{\# edges (ties): $t_U\subseteq U\times U$} & \scriptsize 5,327,041 & \scriptsize 618,970  \\
  \hline
  \scriptsize{Av. degree: ${U}/{t_U}$ } & \scriptsize 37.58 & \scriptsize 67.59  \\
\hline
\hline
  \scriptsize{\# nodes (users + friends): $UF$ } & \scriptsize 1,747,783 & \scriptsize 261,780  \\
  \hline
  \scriptsize{\# edges (ties): $t_{UF}\subseteq UF\times UF$}  & \scriptsize 74,585,447 & \scriptsize 25,293,730  \\
  \hline
  \scriptsize{Av. degree: ${UF}/{t_{UF}}$):} & \scriptsize 42.67 & \scriptsize 96.62  \\
\hline
\hline
  \scriptsize{\begin{tabular}[l]{@{}l@{}} \# nodes (users + friends + \\ fof): $UFF$ \end{tabular}} & \scriptsize 7,954,935 & \scriptsize 1,155,324  \\
\hline
\hline
  \scriptsize{Mean average path length} & \scriptsize 4.152 $\pm$ 0.58 & \scriptsize 3.8 $\pm$ 0.57  \\
  \hline
  \scriptsize{Clustering coefficient \cite{watts1998collective}} & \scriptsize 0.104 & \scriptsize 0.1438  \\
  \hline
  \hline
  \scriptsize{\# locations visited by $U$} & \scriptsize 30,630 & \scriptsize 26,780  \\
  \hline
  \scriptsize{\# check-ins by $U$} & \scriptsize 1,983,772 & \scriptsize 1,164,085 \\
  \hline
  \scriptsize{Av.\# check-ins per user and day} & \scriptsize 0.12 & \scriptsize 1.04  \\
  \hline
  \scriptsize{Av.\# check-ins per location} & \scriptsize 64.77 $\pm$ 436.17 & \scriptsize 43.47 $\pm$ 139.72  \\
  \hline
  \scriptsize{Av.\# check-ins per user} & \scriptsize 13.99 $\pm$ 37.57 & \scriptsize 126.90 $\pm$ 84.28 \\
  \hline
  \scriptsize{Av.\# locations per user} & \scriptsize 8.71 $\pm$ 17.45 & \scriptsize 62.35 $\pm$ 31.99  \\
  \hline
  \scriptsize{Av.\# users per location} & \scriptsize 40.33 $\pm$ 282.80 & \scriptsize 21.36 $\pm$ 71.11  \\
  \hline
  \scriptsize{\begin{tabular}[l]{@{}l@{}} Av.\# check-ins per user \\ and location \end{tabular}} & \scriptsize 1.61 $\pm$ 3.35 & \scriptsize 2.04 $\pm$ 4.74 \\
  \hline
  \scriptsize{Av. degree of repetition} & \scriptsize 0.61 $\pm$ 3.35 & \scriptsize 1.04 $\pm$ 4.74  \\
  \hline
  \scriptsize{Av. user entropy} & \scriptsize 1.15 $\pm$ 0.99 & \scriptsize 3.48 $\pm$ 0.71  \\
  \hline
  \scriptsize{Av. location entropy} & \scriptsize 1.73 $\pm$ 1.60 & \scriptsize 1.66 $\pm$ 1.57  \\
  \hline
\end{tabular}
\end{center}
\setlength{\abovecaptionskip}{0ex}
  \caption[Descriptive statistics of Foursquare data-set]{\footnotesize Descriptive statistics of Foursquare data-set}
  \label{table1}
\end{table}

The clustering coefficients of both social networks with all users and only active users are 0.104 for our dataset and 0.1438 for Facebook respectively. The clustering coefficient is a good indicator of the existence of a real social network among a set of users and their relationships \cite{Watts2004}. In order to assert that the social network induced by the Foursquare users represents a valid social network, we compared its clustering coefficient with the clustering coefficient of a randomly generated social network with the same number of ties and the same average degree using the Poisson random graph method presented in \cite{Newman2003}. The clustering coefficient of the random graph was found to be $0.0002$ on average, which considerably lower compared to the clustering coefficients of the Foursquare social networks. The numbers point to the conclusion that the social network induced by the Foursquare users can be assumed to indeed represent a valid social network.

The check-in statistics for the active users show that on average the locations are visited by many users (21.36), and each user visits many locations (62.35) with a very low frequency (2.04) with an average degree of repetition of (1.04), which means most of the locations are publicly accessible. We calculated entropy values for each user to visit location, and each location to be visited by users. The average user and location entropy for the active are found to be (3.48) and (1.66) respectively, which substantiates the finding that the locations are rather publicly accessible.
\section{Correlations between Social Proximity and Mobile Homophily}
\label{correlations}
\subsection{Measures of Social Cohesion}
Social cohesion can be calculated using different approaches depending on the social proximity measurements they rely on such as Neighborhood-, Distance-, Density- and Cluster-based measurements. We used three neighborhood-based measurements:

\emph{Common Neighbors ($CN$)}: The number of common friends between two users. Social cohesion between two users is higher, the higher the number of their common friends (\cite{barabasi2011}).

\emph{Adamic-Adar (AA)}: The measurement common number does differentiate between the common neighbors of two users. A user with a high degree (a popular user with thousands of friends) is a potential common neighbor of ($n(n-1)/2$) pairs of users. Adamic \& Adair therefore use a normalized version of common neighbor $CN$. It penalizes the contribution of each neighbor $u_k \in CN(u_i, u_j)$ by the inverse logarithm of their degree \cite{adamic, barabasi2011}.

\emph{Jaccard Coefficient (Jacc)}: Jaccard coefficient sets the number of common neighbors of two users in relation to the total number of friends of both users. The higher the ratio of common neighbors between two users compared to their total friends, the higher the social cohesion (\cite{barabasi2011}).

Additionally we use one density based measurement, namely \emph{Degree of Cliquishness (DoC)}. DoC quantifies to which extent the friends of two users build a cohesive group. The social cohesion between two users is higher if their friends are interconnected more closely.

\subsection{Measures of Mobile Homophily}

Mobile homophily (proximity) refers to the extent of overlap between the movements of two individuals \cite{barabasi2011}. We calculate mobile homophily between two individuals using following measurements within the emphasize of spatial (first three) and spatial-temporal (last) overlap:

\emph{Spatial Co-location Count (Col):} a \emph{spatial co-location} is a location visited by two users, but not necessarily at the same time. $Col$ simply counts the cases in which two users visit the same location within a time frame of one week (\autoref{Col1we}).

\begin{equation}\label{Col1we}
    Col(u_i,u_j) = \Sigma_{l} \Sigma_{s^{(u_i)}_l \in H_{u_i}} \Sigma_{s^{(u_j)}_l \in H_{u_j}} \Theta (W - |T_{s^{(u_i)}_l} - T_{s^{(u_j)}_l}|)
\end{equation}

$H_{u_i}$ is the set of visits of user $u_i$ and $\Theta$ is the Heaviside step function for two visits $s^{(u_i)}_l$ and $s^{(u_j)}_l$ of both users $u_i$ and $u_j$ to the same location $l$ within a time frame of one week $W$ (\cite{barabasi2011}).

\emph{Spatial Co-location Rate (SCol):} The probability that two users $u_i$ and $u_j$ visit the same location within a period of one week:

      \begin{equation}\label{SCol}
        SCol(u_i, u_j) = \sum_{l_k \in L} p^{(u_i)}(l_k, t) * p^{(u_j)}(l_k, t)
      \end{equation}

where $p^{(u_i)}(l_k, t)$ is the probability of user $u_i$ to follow user $u_j$ to location $l_k$ within a time frame of one week ($t=1$ week).$SCol$ assumes the visits of both users to occur independently (\cite{barabasi2011}), thus this measurement must show no correlation to the social proximity between the two users, otherwise the assumption is rejected and the visit of the one user $u_i$ depends on the visit of the other user $u_j$.

\emph{Spatial Cosine Similarity (SCos):} This measurement refers to the degree of spatial overlap between the trajectories of two users disregarding the time of the visits, i.e. the co-presence of both users at the same location (\cite{barabasi2011}).

\emph{Social Situation ($\mathfrak{s}$) Rate:} is a measurements within the emphasize of spatial-temporal overlap. These measurements guarantee that the two users are present during the same time at the same location, meaning that both users are co-present (\cite{Lehmann2010}). We assume that at least two users  visiting a location within a time frame of one hour to be involved involved in a social situation. The mass $\mathfrak{s}(u_i,u_j)$ is calculated by counting the number of cases where two users $u_i$ and $u_j$ visit the same location within a time frame $\Delta t$, normalized by the total number of times when both users were observed within this time frame $\Delta t$ of one hour (\autoref{Col1hr}).

\begin{equation}\label{Col1hr}
\mathfrak{s}(u_i,u_j) = \frac{\Sigma_{l} \Sigma_{s^{(u_i)}_l \in H_{u_i}} \Sigma_{s^{(u_j)}_l \in H_{u_j}} \Theta (\Delta t - |T_{s^{(u_i)}_l} - T_{s^{(u_j)}_l}|)}
                        {\Sigma_{s^{(u_i)} \in H_{u_i}} \Sigma_{s^{(u_j)} \in H_{u_j}} \Theta (\Delta t - |T_{s^{(u_i)}} - T_{s^{(u_j)}}|)}
\end{equation}

where $H_{u_i}$ is the set of visits of user $u_i$, $s^{(u_i)}$ is any visit of user $u_i$, $\Theta$ is the Heaviside step function for two simultaneous visits $s^{(u_i)}_l$ and $s^{(u_j)}_l$ of both users $u_i$ and $u_j$ to occur within the time frame $\Delta t$ at the same location $l$.

Additionally we use a weighted version of the above measurements using the following weighting factors:

\emph{Location density:} represents the density of other locations in the vicinity of a location $l_i$. Visits of two users to the same location $l_i$ are weighted higher, the higher the density of other locations is in the vicinity of location $l_i$. The pragmatics behind this weighting schema is based on the assumption that in highly populated areas (which is assumed to have high density of specific public locations) a co-location event is less likely than in a sparsely populated area where the few specific locations act as focal points attracting people. 

\emph{Distance From Home Location:} 
If users live within a short distance from each other, the probability of being co-located by chance only is much higher then compared to a situation where their center of life locations are located farther apart. 
We thus weight visits of two users to a location with the logarithm of the distance between their home locations. Note: Due to the lack of information, we assume the center of the region with high check-in density as the home location of a user (\cite{Volkovich2012, Noulas2012, Noulas2011, scellato2011c}).

\emph{Location Population $\rho(l_k)$:} People usually tend to purposefully meet important friends at locations with low population such as their homes or small restaurants/bars in the vicinity of their homes, rather than at locations with high population such as subway station. Each visit to a location $l_k$ is weighted as inversely proportional to the log of the population size $|\rho(l_k)|$ at that location \cite{barabasi2011}. 

\emph{Location Entropy:} The significance of user meetings can be assumed be higher at low entropic locations than at high entropic locations which can be rather be assumed to be public locations that many people frequently visit (such as large subway stations). Thus another weighting schema inversely proportinal to the location entropy is introduced. 

\begin{equation}\label{shannon}
  H(l) = - \sum_{i} p(u_i,l)\ \ln p(u_i,l)
\end{equation}

%
%
\subsection{Correlations}
Mobile homophily refers to the tendency of similar individuals to be interested in the same locations \cite{barabasi2011}. An analysis of correlation between social and mobile homophily was conducted in order to assert mobile homophily among social connected individuals. \autoref{commonLocCheckin} contains first evidences for mobile homophily as it shows higher mobile homophily among friends compared to random pairs. Friends share on average (4.29) locations, whereas random pairs share only (1.61) locations on average. Further, friends are involved on average in (5.74) social situations, whereas the corresponding number for random pairs is only (0.14).

\begin{table}[htb]
\begin{center}
\begin{tabular}[c]{|l|l|l|}
\hline
\scriptsize{\textbf{Type}} & \scriptsize{\textbf{$\phi$ Common locations}} & \scriptsize{\textbf{$\phi$ Social situations}} \\
\hline \hline
  \scriptsize{\emph{Friends:}} & \scriptsize 4.29 & \scriptsize 5.74 \\
  \hline
  \scriptsize{\emph{Random pairs}} & \scriptsize 1.61 & \scriptsize 0.14 \\
  \hline
\end{tabular}
\end{center}
\setlength{\abovecaptionskip}{1ex}
  \caption[Friends have on average 2.5 times more common locations than random pairs of users and are involved in about $\approx$ 40 times more social situations within 1 hour.]{\footnotesize Friends have on average 2.5 times more common locations than random pairs of users and are involved in about $\approx$ 40 times more social situations within 1 hour.}
  \label{commonLocCheckin}
\end{table}

Furthermore, a correlation analysis between mobile homophily and social cohesion was conducted. We use 100\,000 randomly chosen sample pairs of users from the social network $G$ with the whole population. We sample pairs of users using a simple random sampling with replacement method. The users in a sample pais are not necessarily socially connected. The hypothesis is that mobile mobile homophily correlated with social cohesion. We refer to certain intervals of the correlation coefficient with the following equivalences: $\geq  0.7$ corresponds to a very strong correlation, $[0.4, 0.7\rceil$ corresponds to a strong correlation , $[0.1, 0.4\rceil$ corresponds to a moderate correlation, $\leq 0.1$ corresponds to weak or non correlation.

\begin{table}[htb]
\begin{center}
  \begin{tabular}{|l|c|c|c|c|}
\hline
 & CN & AA & DoC & Jacc  \\

\hline
\emph{\tiny Scos:} &	-0.056	&	0.008	&	0.16	&	0.004	\\

\hline
\emph{\tiny SCos-User:} &	-0.058	&	0.008	&	0.15	&	-0.017	\\

\hline
\emph{\tiny SCos-Dens:} &	-0.052	&	0.013	&	0.155	&	0.013	\\

\hline
\emph{\tiny SCos-Dist:} &	-0.037	&	0.014	&	0.155	&	0.024	\\

\hline
\emph{\tiny SCos-Entr:} &	-0.048	&	0.008	&	0.155	&	0.004	\\

\hline
\emph{\tiny Scol:} &	-0.171	&	0.021	&	0.131	&	0.016	\\

\hline
\emph{\tiny Col:} &	-0.207	&	0.023	&	0.131	&	-0.007	\\

\hline
\emph{\tiny $\mathfrak{s}$:} &	-0.066	&	0.043	&	0.112	&	0.021	\\

\hline
\emph{\tiny $\mathfrak{s}$-Dens:} &	-0.027	&	0.048	&	0.105	&	0.022	\\

\hline
\emph{\tiny $\mathfrak{s}$-Dist:} &	-0.01	&	0.07	&	0.084	&	0.014	\\

\hline
\emph{\tiny $\mathfrak{s}$-User:} &	-0.029	&	0.043	&	0.109	&	0.014	\\

\hline
\emph{\tiny $\mathfrak{s}$-Entr:} &	-0.029	&	0.048	&	0.108	&	0.019	\\

\hline
  \end{tabular}
\end{center}
\setlength{\abovecaptionskip}{0ex}
  \caption[Pearson's correlation coefficient $r$ between mobile homophily and network proximity for 100\,000 randomly chosen pairs of users (setting $\Delta t = 1$ hour for spatial-temporal overlap).]{\footnotesize
Pearson's correlation coefficient $r$ between mobile homophily and network proximity for 100\,000 randomly chosen pairs of users (setting $\Delta t = 1$ hour for spatial-temporal overlap).}
  \label{correlationAnalysisAll}
\end{table}

The result of the correlation analysis is shown in \autoref{correlationAnalysisAll}. A slight correlation is noticeable for DoC and all mobile homophily measurements, the remaining measurements show no correlation. The anti-correlation can be explained by measurement artifacts of the data-set, because the location data is limited to a city (San Francisco), whereas the social network is global and contains users from the whole world. A reliable correlation might not be determined for many pairs of users due to the lack of location data.

\subsubsection{Impacts of Propinquity on Social Cohesion}

Tobler's first law of geography states that "everything is related to everything else, but near things are more related than distant things" \cite{Tobler1970}. \emph{Propinquity} refers to the tendency of individuals to have their ties with other in their geographical vicinity \cite{Kadushin2012}. 
For example, two users living in the same building have a higher propinquity than two users living in different buildings \cite{Festinger1950}. We constrain the social network to the active users and their friends and friends of friends (FoF) from San Francisco (home city) in order to investigate the effects of geographical constraints on the social network, in accordance to Tobler's statement and the propinquity effect. We refer to the induced graph by $G_{HC}$.


\begin{table}[htb]
\begin{center}
\begin{tabular}[c]{|l|l|l|l|}
\hline
\scriptsize \textbf{Type} & \scriptsize{\textbf{\begin{tabular}[l]{@{}l@{}} Clustering \\ Coefficient \end{tabular}}} & \scriptsize{\begin{tabular}[l]{@{}l@{}} \textbf{$\phi$ Mean average} \\ \textbf{path length:} \end{tabular}} & \scriptsize{\textbf{\begin{tabular}[l]{@{}l@{}} Av. \\ Degree \end{tabular}}} \\
\hline \hline
  \scriptsize{\emph{All:}} & \scriptsize 0.104 & \scriptsize 4.152 \scriptsize $\pm$ \scriptsize 0.58 & \scriptsize 42.67 \\
  \hline
  \scriptsize{\emph{Active users:}} & \scriptsize 0.1438 & \scriptsize 3.8 \scriptsize $\pm$ \scriptsize 0.57 & \scriptsize 96.62\\
  \hline
  \scriptsize{\emph{Active home city}} & \scriptsize 0.38118 & \scriptsize 3.6745 \scriptsize $\pm$ \scriptsize 0.55 & \scriptsize - \\
  \hline
\end{tabular}
\end{center}
\setlength{\abovecaptionskip}{0ex}
  \caption[A comparison between the results of both clustering coefficients and average shortest path for three different graphs induced by all users, active users and users from San Francisco respectively.]{\footnotesize
A comparison between the results of both clustering coefficients and average shortest path for three different graphs induced by all users, active users and users from San Francisco respectively.}
  \label{globalLocalClusteringCoefficient}
\end{table}

The induced social network has a higher average degree $67$ compared to all users $37$. The clustering coefficient of $G_{HC}$ is found to be $C_{G_{HC}} \approx 0.38$, which is significantly higher than the clustering coefficients $C_{G} \approx 0.104$ and $C_{G_{AU}} \approx 0.1438$. The average shortest path increases from $4.152 \pm 0.58$ for $G$ to $3.6745 \pm 0.55$ for $G_{HC}$. The significance of the change in the average shortest path is confirmed by a two-sided unpaired t-test $ p(\epsilon)=1.09 * 10^{-294}$). The results show indeed that geographically close people form a closer knit network. A new correlation analysis using random pairs of users from $G_{HC}$ substantiates the above findings. Table \ref{correlationAnalysisHomecity} shows how the correlation between mobile homophily and network proximity significantly increases. The significance of the changes in CN, AA, Jacc and DoC is confirmed by 3 two-sided unpaired t-tests (with $p(\epsilon)$ values $0.000601, 0.0122 \& 0.000124$ respectively).

\begin{table}[htb]
\begin{center}
  \begin{tabular}{|l|c|c|c|c|}
\hline

 & CN & AA & DoC & Jacc  \\

\hline
\emph{\tiny Scos:} &	0.417	&	0.225	&	0.168	&	0.315	\\

\hline
\emph{\tiny SCos-User:} &	0.24	&	0.185	&	0.184	&	0.327	\\

\hline
\emph{\tiny SCos-Dens:} &	0.418	&	0.221	&	0.169	&	0.32	\\

\hline
\emph{\tiny SCos-Dist:} &	0.389	&	0.154	&	0.153	&	0.273	\\

\hline
\emph{\tiny SCos-Entr:} &	0.267	&	0.192	&	0.175	&	0.313	\\

\hline
\emph{\tiny Scol:} &	0.141	&	0.101	&	0.188	&	0.25	\\

\hline
\emph{\tiny Col:} &	0.309	&	0.122	&	0.188	&	0.234	\\

\hline
\emph{\tiny $\mathfrak{s}$:} &	0.326	&	0.242	&	0.151	&	0.25	\\

\hline
\emph{\tiny $\mathfrak{s}$-Dens:} &	0.309	&	0.281	&	0.16	&	0.341	\\

\hline
\emph{\tiny $\mathfrak{s}$-Dist:} &	0.342	&	0.272	&	0.169	&	0.353	\\

\hline
\emph{\tiny $\mathfrak{s}$-User:} &	0.215	&	0.231	&	0.178	&	0.349	\\

\hline
\emph{\tiny $\mathfrak{s}$-Entr:} &	0.28	&	0.261	&	0.176	&	0.35	\\

\hline
\emph{\tiny $\mathfrak{s}$-Extra Role:} &	0.052	&	0.163	&	0.145	&	0.351	\\
\hline
  \end{tabular}
\end{center}
\setlength{\abovecaptionskip}{1ex}
  \caption{\footnotesize
Pearson's correlation coefficient $r$ between mobile homophily and network proximity for 100,000 randomly chosen pairs of users from $G_{HC}$ (setting $\Delta t = 1$ hour for spatial-temporal overlap).}
  \label{correlationAnalysisHomecity}
\end{table}

\subsubsection{Impacts of Cohesive Subgroups on Mobile Homophily}\label{corrCohesion}

People typically, maintain their social relationships on different scales due to cognitive, emotional, spatial, and temporal limits. Therefore, they interact mainly with a small group of their acquaintances (\cite{Wilson2009, Jiang2010} as cited by \cite{Kaltenbrunner2012}), who together form a cohesive subgroup with a higher probability. From the behavioral-perspective, members of a cohesive subgroup share information and have homogeneity of interests and beliefs\cite{Balabhaskar2009}, therefore they exhibit a high similarity in their behavior including mobile homophily. Different types of cohesive subgroups exist from the perspective of social network analysis such as cliques and n-plexes. The main property of cliques is the completeness of the ties inside the groups, where all members are connected to each other. N-plexes relax the completeness constraint of cliques by allowing for up to $N$ missing connections for each group member.

We use the algorithm proposed by \cite{Tsukiyama1977} for enumerating maximal cliques based on a binary tree with $n$ levels. We stopped the enumeration after having detected 8700 cliques with a minimum size of 3. The largest clique contains 17 users. The completeness requirement of cliques makes them impractical for the detection of real-life cohesive subgroups, therefore we slightly relax the cliques to 2-plexes in order not to change the clique properties significantly. We detect $29591$ 2-plexes with a minimum size of $3$, the plexes contain on average $7.79 \pm 2.41$ vertices. The largest 2-plex contains $20$ vertices, which is in rough agreement with the human social perception limit in \cite{Fischer1997}.


We repeated our correlation analysis by sampling 100\,000 pairs of users, where each pair is chosen from the same 2-plex. Table \ref{correlationAnalysis2plex} compares the average values of all mobile homophily and social cohesion measurements with the corresponding values of the previous correlation analysis between pairs of users from the same home city. All social cohesion measurements are as much as a factor of $50$ and the mobile homophily measurements as much as a factor of $23$ higher. The results show a significantly stronger interdependence between spatial and social cohesion among the members of cohesive subgroups.

\begin{table}[htb]
\begin{center}
  \begin{tabular}{|l|c|c|}
\hline

 & Home city  & 2-plex   \\

\hline
\emph{\tiny Scos:} &	0.013	$\pm$	0.012	&	0.11	$\pm$	0.101	\\

\hline
\emph{\tiny WSCos-Dens:} &	0.014	$\pm$	0.012	&	0.111	$\pm$	0.101	\\

\hline
\emph{\tiny WSCos-Dist:} &	0.008	$\pm$	0.007	&	0.055	$\pm$	0.051	\\

\hline
\emph{\tiny WSCos-User:} &	0.008	$\pm$	0.006	&	0.099	$\pm$	0.075	\\

\hline
\emph{\tiny WSCos-Entr:} &	0.01	$\pm$	0.008	&	0.107	$\pm$	0.09	\\

\hline
\emph{\tiny Scol:} &	0.001	$\pm$	0.001	&	0.009	$\pm$	0.007	\\

\hline
\emph{\tiny Col:} &	1.626	$\pm$	1.27	&	40.006	$\pm$	29.176	\\

\hline
\emph{\tiny $\mathfrak{s}$:} &	0.058	$\pm$	0.03	&	4.538	$\pm$	3.284	\\

\hline
\emph{\tiny $\mathfrak{s}$-Dens:} &	0.006	$\pm$	0.003	&	0.158	$\pm$	0.124	\\

\hline
\emph{\tiny $\mathfrak{s}$-Dist:} &	0.003	$\pm$	0.001	&	0.054	$\pm$	0.042	\\

\hline
\emph{\tiny $\mathfrak{s}$-User:} &	0.002	$\pm$	0.001	&	0.044	$\pm$	0.035	\\

\hline
\emph{\tiny $\mathfrak{s}$-Entr:} &	0.004	$\pm$	0.002	&	0.131	$\pm$	0.103	\\

\hline
\emph{\tiny $\mathfrak{s}$-Extra Role:} &	0.002	$\pm$	0.001	&	0.054	$\pm$	0.037	\\

\hline

\rowcolor[RGB]{151,217,255}
\emph{\tiny CN:} &	0.193	$\pm$	0.103	&	14.842	$\pm$	17.881	\\

\hline

\rowcolor[RGB]{151,217,255}
\emph{\tiny AA:} &	0.102	$\pm$	0.054	&	7.804	$\pm$	10.526	\\

\hline

\rowcolor[RGB]{151,217,255}
\emph{\tiny Jacc:} &	0.005	$\pm$	0.002	&	0.059	$\pm$	0.112	\\

\hline

\rowcolor[RGB]{151,217,255}
\emph{\tiny DoC:} &	0.002	$\pm$	0.001	&	0.082	$\pm$	0.081	\\

\hline
  \end{tabular}
\end{center}
\setlength{\abovecaptionskip}{0ex}
  \caption{\footnotesize The average values for network proximity and mobile homophily measurements for 100\,000 random pairs chosen from 2-plexes and 100\,000 random pairs chosen from $G_{HC}$.}
  \label{correlationAnalysis2plex}
\end{table}

Cohesion is a measure that allows the identification of more important subgroups among the set of all subgroups \cite{Kosub2004}. A more cohesive subgroup has more ties inside and fewer outside the subgroup. Given a cohesive subgroup and the adjacency matrix $A$, $\mathcal{C}(U)$ measures the degree of cohesion in a cohesive subgroup (\autoref{measureOfCohesion} \cite{Wasserman1994} ~Chapter 7):

\begin{equation}\label{measureOfCohesion}
  \mathcal{C}(U) = \frac{\frac{\sum_{v_i \in U}\sum_{v_j \in U} A_{ij}}{|U|(|U|-1)}}{\frac{\sum_{v_i \in U}\sum_{v_j \notin U} A_{ij}}{|U|(|V-U| -1)}}
\end{equation}

We use this measure ito investigate the impacts of groups cohesion on mobile homophily. \autoref{correlation2plexCohesionLoc} shows the results of a correlation analysis between the cohesion measurement according to \autoref{measureOfCohesion} and mobile homophily. Mobile homophily measurements based on spatial overlap (asynchronous) show a correlation coefficient varying between $r=0.22, \rho=0.23$ for co-locations within one week and $r=0.51, \rho=0.65$ for $WSCos-Entr$. Mobile homophily measurements based on spatial-temporal (synchronous) overlap show higher correlation coefficient values, for example for social situation rates ($\mathfrak{s}$) the values vary between $r=0.48, \rho=0.63$. The weighting factors used all led to higher correlation coefficients, with "distance from home" $\mathfrak{s}-Dist$ proving to be the best weighting factor (ranging between $r=0.66, \rho=0.81$), which confirms that people are rather willing to cover greater distances, mostly in order to meet close/important friends.

\begin{table}[htb]
\begin{center}
  \begin{tabular}{|l|c|c|c|}
\hline

 & $r$ & $\rho$ & $p(\epsilon)$  \\

\hline
\emph{\tiny Scos:} &	0.45	&	0.6	&	0	\\

\hline
\emph{\tiny WSCos-Dens:} &	0.43	&	0.6	&	0	\\

\hline
\emph{\tiny WSCos-Dist:} &	0.5	&	0.63	&	0	\\

\hline
\emph{\tiny WSCos-User:} &	0.46	&	0.67	&	0	\\

\hline
\emph{\tiny WSCos-Entr:} &	0.51	&	0.65	&	0	\\

\hline
\emph{\tiny Scol:} &	0.15	&	0.43	&	0.0006	\\

\hline
\emph{\tiny Col:} &	0.22	&	0.23	&	0.0562	\\

\hline
\emph{\tiny $\mathfrak{s}$:} &	0.48	&	0.63	&	0	\\

\hline
\emph{\tiny $\mathfrak{s}$-Dens:} &	0.57	&	0.73	&	0	\\

\hline
\emph{\tiny $\mathfrak{s}$-Dist:} &	0.66	&	0.81	&	0	\\

\hline
\emph{\tiny $\mathfrak{s}$-User:} &	0.6	&	0.78	&	0	\\

\hline
\emph{\tiny $\mathfrak{s}$-Entr:} &	0.58	&	0.79	&	0	\\

\hline
\emph{\tiny $\mathfrak{s}$-Extra Role:} &	0.32	&	0.75	&	0	\\
\hline






  \end{tabular}
\end{center}
\setlength{\abovecaptionskip}{0ex}
  \caption{\footnotesize
The correlation between 2-plex cohesion measurement calculated according to equation \ref{measureOfCohesion} and all mobile homophily measurements. The last column contains the p-value of the corresponding Spearman's correlation coefficient. The p-value indicates the probability that social proximity and mobile homophily have no relationship (page \pageref{lst:spearmanCorCoef}). The correlation between 2-plex cohesion measurement calculated according to equation \ref{measureOfCohesion} and all social network measurements}
  \label{correlation2plexCohesionLoc}
\end{table}


\begin{figure}[htb]
\begin{center}
 \includegraphics[width=0.5\textwidth]{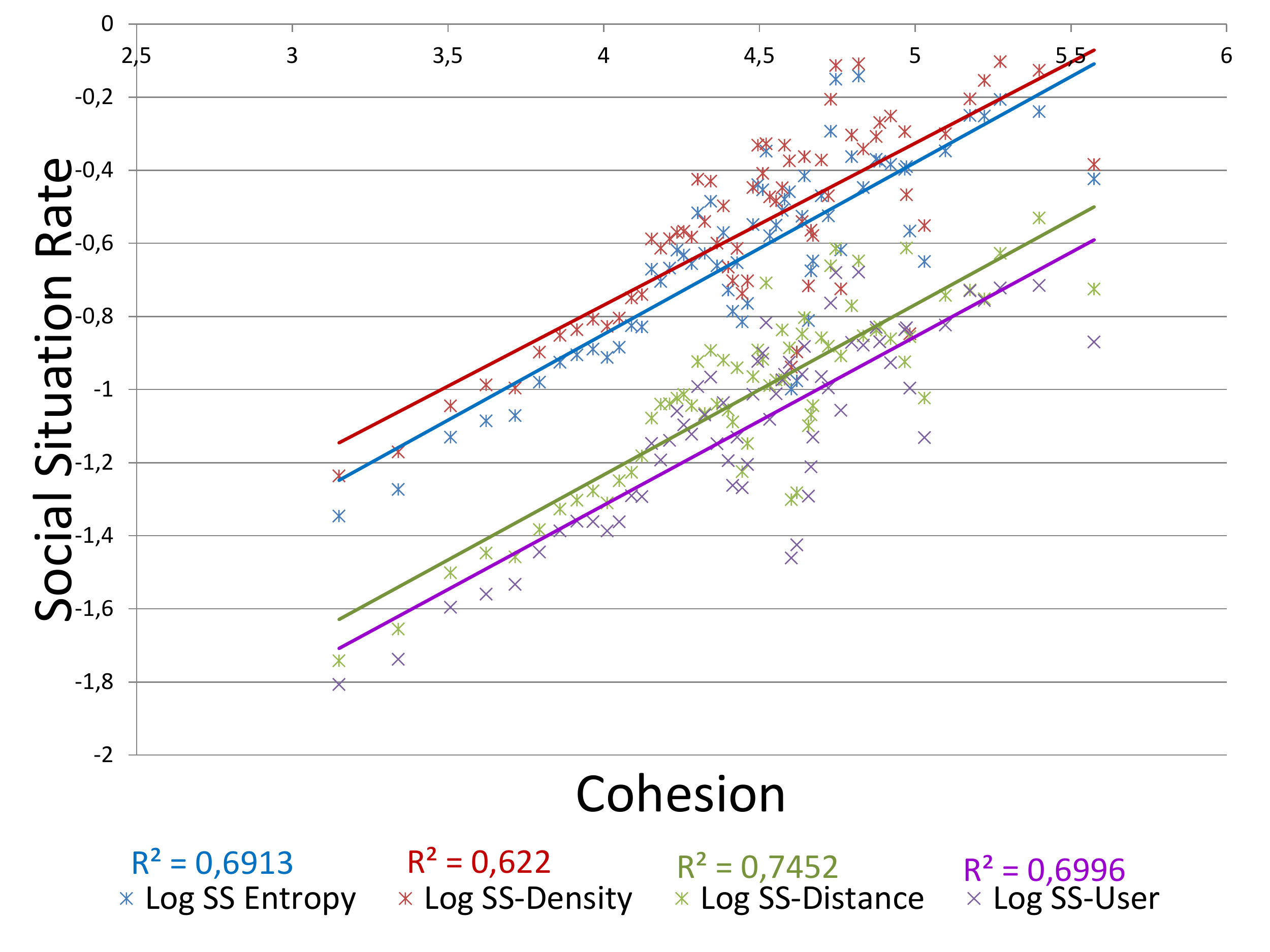}
\end{center}
\setlength{\abovecaptionskip}{0ex}
\caption{\label{2plexCohesionSSRate}
\footnotesize The correlation between the cohesion measurement according to equation \ref{measureOfCohesion} and different weighted social situation rates.}
\end{figure}


The relationship between group cohesion and mobile homophily is shown on a log-log plots on figure \ref{2plexCohesionSSRate} for measurements based on spatial overlap. The results show a more monotonic relationship than linear relationship, which explains the higher coefficient values for Spearman's rank correlation coefficient. Moreover, the log-log plot for the spatial-temporal overlap shows a higher coefficient of determination $R^2=0.74$ than the square of the correlation coefficients, which is an evidence for the non-linear nature of the relationship. We noticed a similar relationship between group cohesion and mobile homophily measurements based on spatial-temporal overlap.

The correlation analysis shows reciprocal impacts between mobile homophily and social cohesion. People find others in their geographical vicinity more attractive for a social relationship on one hand side, on the other hand they share more locations with their closest friends (represented by the members of their cohesive subgroups).

\section{Social Relations and Next Location Prediction}
\label{socialPrediction}
\subsection{Related Work}

The correlation between mobile homophily and social cohesion shows a statistical dependence between two events rather than a relationship of cause and effect, i.e. the occurrence of one event causes the occurrence of the second event. The order of the occurrence of the events is significant in causation in contrast to correlation correlation analysis where it is of little importance, because causation requires the cause to precede or coincide with the consequence.

In general, three types of causation effect can be distinguished, namely necessary (a cloud is necessary for rainfall), sufficient (wind is a sufficient cause for the rustling of the trees) and contributory.  According to the definition of \cite{Riegelman1979} for contributory cause, and according to the INUS condition for contributory causes in \cite{Mackie1974}, the influence of the mobility of users on the mobility of their neighbors is a contributory cause. A visit(cause) of a user(influencer) to a location, can cause (contribute to) a visit(effect) of friends(followers) to the same location, either the cause precedes the effect(asynchronous influence, i.e. the influencer is absent during the visit of the friends) or it coincides with it (synchronous influence, i.e. the influencer accompanies the follower during their visit).

The influence of social networks on predicting individual mobility has been investigated by \cite{Musolesi2004, Sadilek2012, weiPan2012, hackney2006, pentland09a, pentland09b, eunjoon2011, jieTang2009}.

\cite{eunjoon2011} proposed two mobility models, namely Periodic Mobility Model (PMM) and Periodic \& Social Mobility Model (PSMM) for predicting the locations of an individual. PMM models the mobility of a user as a time-variant stochastic process, where the temporal dynamics of human mobility are captured based on a day-specific periodic transition model. PMM uses a mixed Gaussian distribution centered around two states for predicting the future location of the user, namely home \& work states, PSMM considers a third social state for evening and weekend activities. Using PSMM improve the average distance error between the predicted and the exact location of the user from 2.9\% to 2.7\%, which corresponds to an improvement by $10\%$.

Random Utility Decision Models (RUM) (or discrete choice models) are statistical procedures that predict the choice of a user among alternatives based on a utility function \cite{Mcfadden01economicchoices}, which can take many factors into account. \cite{hackney2006} proposed a work investigating the interdependence between social network and travel behavior based on RUM. It models a user's utility for a location $j$ based on both the travel cost from a start location to a destination location $j$ and the social influence of the destination location. The social influence of a location is calculated based on factors such as the number of friends at the location and/or number of friends-of-friends.

\cite{jieTang2009} has analyzed topical influence and proposed an influence model called Topical Affinity Propagation (TAP). For each topic they determined a set of representative members of a social network and investigated the influence of a member on their friends. The proposed Topical Factor Graph (TFG) incorporates both user-specific topic distribution and network structure into one probabilistic model. TAP learns the model parameters using the sum-product algorithm described in \cite{sumProduct2001}.

Mobility models based on Dynamic Bayesian Networks DBN has been proposed by \cite{Musolesi2004, sadilek2012, weiPan2012}. The influence of groups on individual human movement has been investigated by \cite{Musolesi2004}. The authors observed a strong influence of groups on individuals moving in and between these groups. An individual either moves independently from groups, or within the sphere of influence of a geographical group at any point in time. An individual joins a group based on the attractiveness (depending on interaction of the user with group members) of the group and the difficulty of reaching that group. \cite{Sadilek2012} has proposed a mobility model taking into account both temporal and social dependencies based on a DBN approach. The model is evaluated using publicly available GPS-tagged tweets from two different areas (LA and NY). \cite{weiPan2012} have proposed a general social influence model that can be applied to any interaction network in any social system (including social networks like Foursquare, Facebook, etc.). Their general social influence model is based on a simple mixture approach with fewer parameters than the Hidden Markov models called the dynamical influence model. They proposed two models, a model with a static tie strength matrix, and a more dynamic extended model that uses a set of different tie strength matrices for capturing dynamic changes over time. The authors use a switching latent state variable which controls the current tie strength matrix to be used. The dynamic influence model captures how the state of one user is influences by the state of their neighbour.

\subsection{Social Context and VOMM-Based Location Prediction}

We have introduced in a previous work \cite{grohBapierreTheiner} a mobility model using an adapted general-purpose algorithm called Prediction by Partial Matching (PPM) \cite{begleiter2004}, which is based on variable markov models (VOMM). The proposed mobility model considers both spatial and temporal context while predicting the next location of a mobile user \cite{grohBapierreTheiner}. PPM, in contrast to Bayesian Network (BN) models, predicts the value of a random variable based on a subset of random variables of variable size depending of the specific realization of observed variables in the training data called context (s). The size of the context $n=|s|$ represents the order of the model. The variable order of PPM alleviates the negative impacts of missing data and zero-frequency due. VOMM approaches generally use a tree structure to alleviate the problem of transition matrix sparseness.

Given the current context $s$, PPM estimates the probability of a location $q$ to be visited after $s$ by estimating the conditional probability $p(q|s)$. PPM assigns a probability mass $P(escape|s)$ to symbols that does not appear after context $s$, the remaining probability mass $1-P(escape|s)$ is distributed among the symbols appearing after $s$. \autoref{recursion} determines the probability of any symbol $q$ occurring after context $s$ recursively.

\begin{equation}
\label{recursion}
    P(q|s) = \left\{
\begin{array}{l l}
\tilde{P}(q|s) & \mathrm{if} \; q \in \Sigma_s\\
\tilde{P}(escape|s) \; P(q|\mathrm{suf}(s)) &  \mathrm{else}\\
\end{array}
\right .
\end{equation}
where $\Sigma_s$ is the set of symbols appearing after context $s$, $\mathrm{suf}(s)$ denotes the longest suffix of $s$. The probability of any symbol appearing after an empty context $|s|=0$ is $P(q|\varepsilon)=\frac{1}{|\Sigma|}$. Hence, PPM is able to assign a probability mass to any symbol independently of its occurrence in the training sequence, thus PPM does not contravene Cromwell's rule (compare with the rule of succession in general statistics \cite{zabell1989rule}). For symbol $q$ and context $s$, let $C(sq)$ be the counter that counts the occurrences of $sq$, \autoref{eq_pmm1} and \autoref{eq_pmm2} are estimations of both probabilities $\tilde{P}(q|s)$ and $\tilde{P}(escape|s)$ respectively:
\begin{equation}\label{eq_pmm1}
    \tilde{P}(q|s) = \frac{C(sq)}{|\Sigma_s| + \Sigma_{q'\in\Sigma_s}C(sq')}
\end{equation}
\begin{equation}\label{eq_pmm2}
    \tilde{P}(escape|s) = 1 - \sum_{q\in\Sigma_s}\tilde{P}(q|s) = \frac{|\Sigma_s|}{|\Sigma_s| + \Sigma_{q'\in\Sigma_s}C(sq')}
\end{equation}

We integrate the temporal context into the PPM model making use of the inclusion sematic of human kind periods (Week > day > hours), which has the same hierarchical structure as the PPM tree \cite{grohBapierreTheiner}. For example, a spatial context $s$ can appear in temporal contexts of the form $(Working day, Tuesday, 7 pm)$.

\subsubsection{Integration of Social Context}

The Socio-Spatial-Temporal (SOST) PPM is an improvement of the spatial-temporal PPM model that incorporate social influence factors. Generally, we distinguish between two types of social influence factors, namely synchronous specific and general social trend influences. A Synchronous specific influence factor represents the cases when a user and a set of their friends are involved in the same social situation (when they visit a restaurant for example). We refer to the friends from whom the influence originates as influencers. Thus, synchronous social influence has two preconditions, first the user must be currently involved in a social situation, second the availability of location histories of friends. The introduce Synchronous specific influence factors in more details in the next subsection.

General social influence factors represent the general movement patterns and social trends in a user's community, for example the favorite bar or club of their circle of friends, a hip new restaurant in the city or an inexpensive shopping mall, etc. The users in the same circle of friend or the same community share common interests, hobbies, thoughts, beliefs, etc. which precipitate interest in common locations, or similar movement behavior in the same spatial-temporal context. General social influence has only one precondition, namely the availability of location histories of friends.

A precondition of synchronous specific social influence factors in the presence of the users at the same location during a short period of time $\Delta t$. The correlation analysis has a shown a moderate to a strong correlation between mobile homophily and social cohesion when setting $\Delta t$ to one hours, therefore we assume that friends, who are present at a location within one hour to be involved in the same social situation.

The Social context of a visit to a location from the point of view of a user $u_i$ is a tuple $\mathfrak{s} = <U, q, \lambda, t, c>$, where $U$ is the set of users present at location $q$ representing a synchronous specific social influence factor, $\lambda$ represents the set of temporal features (such as work day or weekend, day of week, hours of day etc.) extracted from the time stamp $t$ of the visit and $c$ is a counter that bookkeeps the occurrence of that specific social influence factor. Let user $u_i$ be the user whose next location is going to be predicted and $N(i)$ the neighbors (friends) of $u_i$, then $U$ is a subset of $U \subseteq N(i) \cup u_i$. According to the users present in $U$, the social context of a visit can be categorized into three different classes of synchronous specific social influence factors $\mathfrak{s}$:

\begin{itemize}
  \item \emph{Class I social influence factors -} is a social situation that contains the user $u_i$ and at least one of their neighbors, i.e. $u_i$ is visiting a location with at least one of their friends.
  \item \emph{Class II social influence factors -} is a social situation that contains at least two neighbors of the user $u_i$ without the presence of $u_i$ in the social situation.
  \item \emph{Class III individuals based social influence factors -} Contains single visits of the neighbors without the presence of other users.
\end{itemize}

The inclusion of class II \& III social influence factors injects a vast amount of extra knowledge into the mobility model of an individual. It helps predict locations which have been visited by friends even if the user themselves has never been there before. The prediction of locations where the user has never been before is almost impossible if only the location history of the individual user is available (but may be possible to a limited extent if additional data sources such as their personal calendar are considered).

The probability a user visits a location under consideration of the spatial-temporal context and synchronous specific social influence factors is captured by the conditional probability $P(q|U, s,\lambda)$. We assume that the current social situation $U$ is independent from the spatial context. \autoref{specificInfuence} is an estimation of the the probability mass $P(q|U, s,\lambda)$:

\begin{equation}\label{specificInfuence}
    P(q|U, s, \lambda) = \underbrace{P(q| U, \lambda)}_{\mathrm{Social\ influence}}*\underbrace{P(q|s, \lambda)}_{\mathrm{Individual\ mobility}}
\end{equation}

The right term of the equation represents the probability of visiting a location given both spatial and temporal contexts, which can be estimated from the individual location history as in the previous chapters. The left term represents influences arising from the current social situation and temporal context of the user.

The inclusion of location histories of friends causes an explosion in the number of locations in the alphabet $\Sigma$, thus we manage social influence factors in a separate tree called Socio-Spatial-Temporal (SOST) tree. \autoref{variablesSOSTPPMVOMM} shows the features and their domains, that are used in the SOST PPM VOMM tree.

\begin{table}[htb]
\centering
\begin{tabular}[c]{|l|l|l|}
\hline
\scriptsize \emph{Variables} & \scriptsize \emph{Domain} & \scriptsize \emph{Description} \\
\hline \hline
  \emph{$\Sigma_{loc}$} & \scriptsize{$\{l_1, l_2, ..., l_i\}$} & \scriptsize {\begin{tabular}[l]{@{}l@{}} The set of locations visited \\ by the user \end{tabular}} \\
  \hline
  \emph{$W$} & \scriptsize{$\{Wd, We\}$} & \scriptsize{\begin{tabular}[l]{@{}l@{}} a binary variable representing \\ whether it is a weekend day \\ or a work day \end{tabular}} \\
  \hline
  \emph{D} & \scriptsize{$\{Sun, Mon, ..., Sat\}$} & \scriptsize{\begin{tabular}[l]{@{}l@{}} The day of week \end{tabular}} \\
  \hline
  \emph{$S^{\Delta t}$} & \scriptsize{$\{S_1, S_2, ..., S_j\}$} & \scriptsize{\begin{tabular}[l]{@{}l@{}} The number of time slots \\ calculated by dividing the \\ hours of day by $\Delta t$, \\ setting $\Delta t = 1$ means \\ that each hour of day \\ represents a slot \end{tabular}} \\
  \hline
  \emph{$U$} & \scriptsize{$\{u_1, u_2, ..., u_j\}$} & \scriptsize{\begin{tabular}[l]{@{}l@{}} The set of users present in \\ the current social situation \end{tabular}} \\
  \hline
\end{tabular}
\setlength{\abovecaptionskip}{1ex}
  \caption[The features included in the SOST PPM VOMM model.]
  {\footnotesize The features included in the SOST PPM VOMM model.}
  \label{variablesSOSTPPMVOMM}
\end{table}

\begin{figure}[htb]
\centering
 \includegraphics[width=0.5\textwidth]{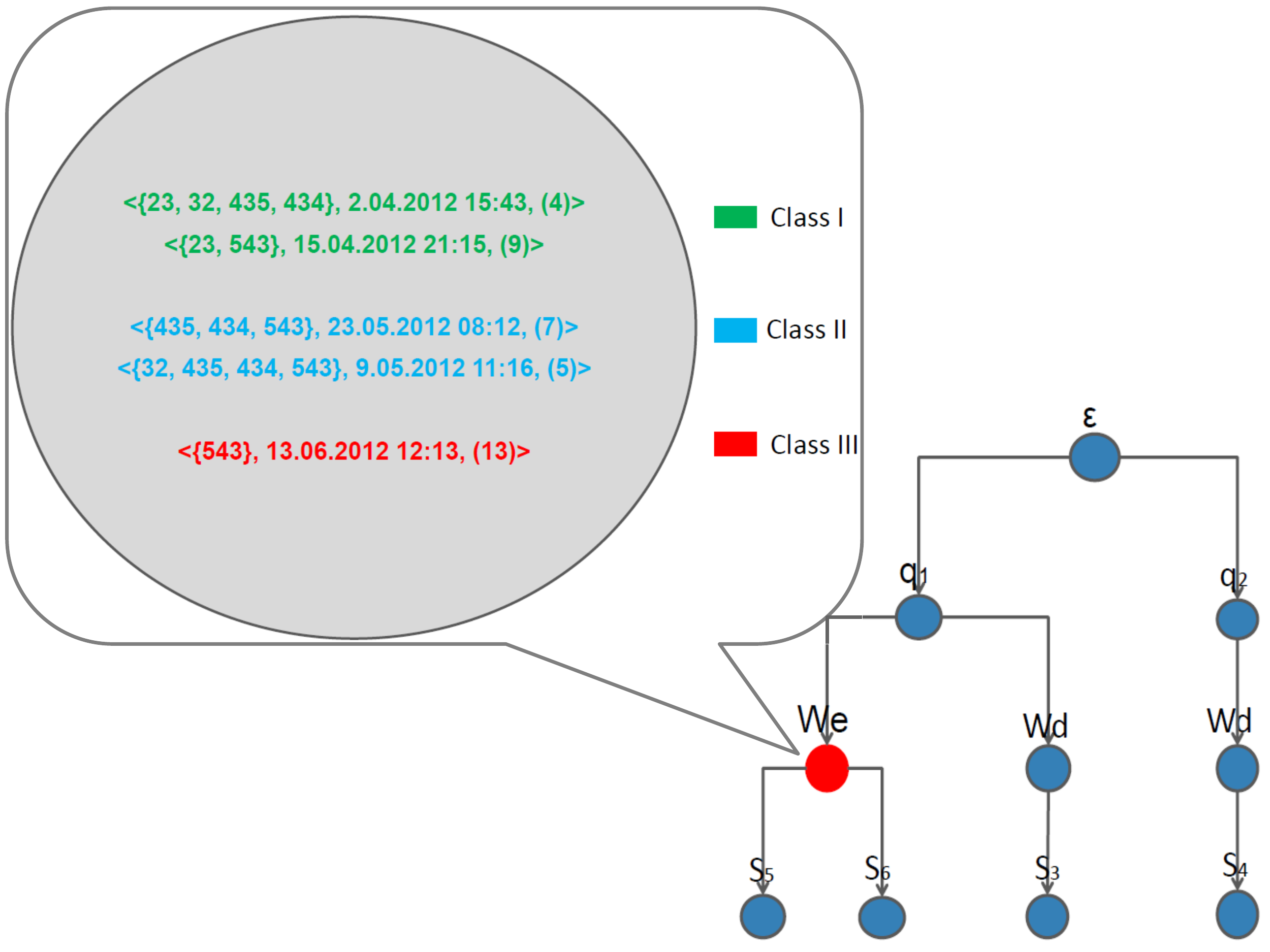}

\setlength{\abovecaptionskip}{0ex}
\caption{\label{socialNode}
\footnotesize An example SOST PPM VOMM tree of a user with ID = 23, the nodes immediately under the root node are labeled with locations, the nodes at deeper levels are labeled with temporal features such as work and weekend days and time slots of day. Unlike the PPM VOMM tree on figure (\ref{vmmTree}), each node in the SOST PPM VOMM tree has multiple tuples for managing the occurrence of social influence factors. The figure zooms into the red node in order to illustrate how SOST PPM VOMM manages the three classes of social influence factors at location $q_1$ on weekend days. Each social influence factor is a tuple consisting of a set of users who build together the social influence factor (the numbers in curly braces represent the IDs of the users), the time stamp of its latest occurrence and a counter (the number in parenthesis) for bookkeeping the number of its occurrences. Social influence factors of different classes are colored differently.}
\end{figure}

The nodes of the SOST tree corresponds to either locations, or temporal features from the location histories of a user $u_i$ and their friends $N(i)$. Each node of SOST tree manages a set of tuples of form $<U, t, c>$ for the social situations in the location histories of the friends at the given spatial-temporal context associated with the node. $U$ represents the users involved in a social situation, $t$ the time stamp of the latest occurrence of $U$ and $c$ is a counter for bookkeeping the occurrence of $U$. Figure \ref{socialNode} shows an example SOST PMM VOMM tree and zooms into a node in order to illustrate how SOST PMM manages different social influence factors.

Upon detecting a social situation influence factor, we traverse the SOST tree to find a corresponding node according to the current location $q$ and temporal context $\lambda$ of each of the social influence factors. We insert a new path if a corresponding node does not yet exist in the tree (necessary for class II \& III social influence factors). We initialize a new counter for a social influence factor that occurs for the first time. The integration of social context into our mobility model simply corresponds to adding new paths to the SOST VOMM PMM tree, incrementing or initializing counters for each different social influence factor. SOST PPM VOMM increments the occurrence of a synchronous specific social influence factor according to \autoref{jaccardDrift}

\begin{equation} \label{jaccardDrift}
	C^t_{U}(q, \lambda) = C^{t_i}_{U}(q,\lambda) (\psi(U, t) + 1)
\end{equation}

where the factor $\psi(U, t)$ represents the degree of drift of the social situation $U$, because Social situations are subject to decay over time. Members of the same social group exhibit similarities in their beliefs, interests, hobbies, goals, activities, emotional needs, feeling of security etc. Group affiliation and aforementioned similarities are subject to decay overtime, thus the influence of previous social situations decreases the longer their last occurrence is in the past. SOST tree uses two different functions for estimating the degree of drift $\psi(U, t)$:

\begin{equation}\label{driftFunction}
  \psi_1(U, t) = (1 - \beta)^{(t - t_{i})/\Delta t}
\end{equation}

\begin{equation}\label{driftFunctionExp}
  \psi_2(U, t) = e^{- \beta (t - t_i)/\Delta t}
\end{equation}

where $\beta$ is a hyper-parameter that controls the degree of drift and $U$ a synchronous specific social influence factor and $t_i < t$ is the time stamp of the last occurrence of $U$. The unit of the factor $t - t_{i}$ is the average stay time $\Delta t$ in hours. If a user spends on average three hours at a location, the social influence factor decays every three hours by a factor of $1 - \beta$ or $e^{- \beta}$.

\subsubsection{Synchronous Specific Social Influence Estimation}

SOST PPM VOMM has to estimate the social influence part $P(q| U, \lambda)$ of equation (\ref{specificInfuence}) in order to predict the future location of a user using synchronous specific social influence factors and both spatial and temporal features. Social situations do not have an order of occurrence like the spatial context, and do not follow an inclusion semantic as the temporal context, therefore the standard escape mechanism of PPM is not applicable. SOST PPM instead uses a similarity measurement based on Jaccard Similarity Coefficient (JSC) \cite{JaccardCoefficient} for comparing two different social influence factors. JSC is defined as the size of the intersection divided by the size of the union of two sets of users involved in two social influence factors (\ref{jaccardCoeff}):

\begin{equation} \label{jaccardCoeff}
	\mathrm{Jacc}(U, \acute{U}) = \frac{\sum_{u_j \in U \cap \acute{U}} \mathfrak{T}(u_i, u_j)}{\sum_{u_k \in U \cup \acute{U}}\mathfrak{T}(u_i, u_k)}
\end{equation}

where $\mathfrak{T}(u_i, u_j)$ is the tie strength between the users $u_i$ and $u_j$. We calculate tie strength between two users using the amount of spatial overlap between their movement histories. Let $N(i)$ be the set of neighbors of user $u_i$, SOST VOMM PPM calculates tie strength based on spatial overlap only between user $u_i$ and a friend from the set $N(i)$ according to equation \ref{tieStrength}:

\begin{equation}\label{tieStrength}
  \mathfrak{T}(u_i, u_j) = \frac{\sum_{l \in L}col_l(u_j)\omega(l)}{\sum_{u_k \in N(i)}\sum_{l \in L}col_l(u_k) \omega(l)}
\end{equation}

$\omega(l)$ is a function that weights visits to location $l$ according to weighting factors such as the distance from home, the density or the entropy of a location.

SOST PPM uses \autoref{modifiedCounter} to determine the amount of influence arising from past synchronous specific social influence factors at a location on the current visit of a user to that location by modifying the counter of the node $\eta$ corresponding to the current spatial-temporal context ($q, \lambda$).

\begin{equation} \label{modifiedCounter}
	C_{\mathfrak{s}=(U,\lambda,q)} = \sum_{\acute{U}}C_{\mathfrak{s}=(\acute{U},\lambda,q)}*Jacc(U, \acute{U})
\end{equation}

where $U$ is set of users involved in the current social situation. The counter $C_{\mathfrak{s}=(U,\lambda,q)}$ for the occurrences of a synchronous specific social influence factor with a set of users $U$ at location $q$ and temporal context $\lambda$. SOST PPM makes use of \autoref{modifiedCounter} for estimating the probability mass $P(q|U, \lambda)$. Let $\varsigma$ be the set containing a node $\eta = (q, \lambda)$ corresponding to a location $q$ and temporal context $\lambda$ and its ancestors, (\autoref{eq_pmmSoc}) estimates the probability mass $P(q|U, \lambda)$:

\begin{equation}\label{eq_pmmSoc}
    P(q|U, \lambda) = \frac{C_{\mathfrak{s}=(U,\eta)}}{|\Sigma_{\varsigma}| + \Sigma_{\eta'\in\Sigma_{\varsigma}}C_{\mathfrak{s}=(U,\eta')}}
\end{equation}

\autoref{eq_pmmSoc2} represents another possibility to estimate the probability mass $P(q|U, \lambda)$ by changing the denominator of equation (\ref{eq_pmmSoc2})

\begin{equation}\label{eq_pmmSoc2}
    P(q|U, \lambda) = \frac{C_{\mathfrak{s}=(U,\eta)}}{\sum_{\acute{U}}C_{\mathfrak{s}=(\acute{U},\eta)}}
\end{equation}

The denominator in \autoref{eq_pmmSoc2} is smaller than the denominator in \autoref{eq_pmmSoc}, because it depends only on the node $\eta$ and not its ancestors. Thus \autoref{eq_pmmSoc2} gives synchronous specific social influence factors more importance than \autoref{eq_pmmSoc}.

\subsubsection{General Social (Trend) Influence Estimation}

A user might be under social influence even without the presence of observable influencers. Social trends in the community of a user (for example a hip bar or restaurant) or asynchronous social influences (for example when friends recommend/share locations to/with each other using other media like e-mail, phone, online social network platforms etc.) are examples of general social influences. The inclusion of general social influence in a mobility model is difficult because influence arises from an unobservable subset of the community (the neighbors $N(i)$) of the user and is transmitted with a time delay ranging between few hours to even few weeks. Therefore, we consider general social influences only in cases where the predicted location has a smaller probability than the probability of an unknown location (the probability $P(escape|s)$ in \autoref{eq_pmm2} after escaping to an empty context $s = \varepsilon$). We use beside SOST PPM an additional (general) VOMM tree ($M'$) in order to integrates the trajectories of all friends. The extended mobility model predicts the future location of a user $u_i$ in two steps. The SOST tree is first used to predict the next location of the user $u_i$. If the probability of the predicted location is greater than the escape probability, it returns this location and terminates, otherwise it switches to the new general VOMM tree $M'$ in order to predict the next location of the user (\autoref{asynchronousModel})

\begin{equation}\label{asynchronousModel}
  x = \left\{
        \begin{array}{ll}
          \arg\max\ (P(q |U, s, \lambda)), & P(x |U, s, \lambda) > P(escape|s) \\
          \arg\max\ P'(q |s, \lambda)_{M'}, & \hbox{else}
        \end{array}
      \right.
\end{equation}

\subsection{Results}

We evaluate the performance of the SOST PPM model using the Foursquare data-set from \autoref{dataset}. We define three different limits of predicability in order to asses the performance of SOST PPM. The users visit on average $|L|=62.35$ locations, the average entropy is found to be $3.48$, which means the average minimum number of location necessary for producing an average entropy of $3.48$ is $e^{3.48} = 32.46$, which corresponds to a \emph{Lower bound of Predictability:} of $1/32.46 = 3\%$ \cite{Song2010}). In almost $38\%$ of the cases the users visit new locations where they have never been. Further, a user in the dataset makes on average $2.04$ check-ins at a location. The average check-ins for the remaining $62\%$ locations is found to be $0.38 + x * 0.62 = 2.04, x=2.68$. A mobility model needs at least one of the 2.68 locations for learning, thus a mobility model can at the most achieve an accuracy of 1.68/2.68 = 63\% for 62\% of the check-ins, which corresponds to an \emph{Upper bound of Predictability} of $0.63 * 0.62 = 39 \%$. Finally, we make use of Fano's inequality (\cite{fano1961} as cited by \cite{Song2010}) for calculating the maximum predictability $\Pi^{max}$ based on the entropy and the number of locations visited for each user. The average value of $\Pi^{max}$ over all users is found to be less than $29\%$. Using the average entropy $\mathfrak{E} = 3.48$ and the average number of locations visited by the users $62.35$ of all users, the average predictability increases to $< 31\%$.

The users move in almost $38\%$ to new (not yet seen in their location histories) locations, three-quarter of these locations are previously visited by friends, thus the amount of new locations reduces to $9.8\%$ in the circle of friend of the users. Further, a check-in of a user to a location is followed in $13\%$ of the cases by a check-in of a friend to that location within one hour. Furthermore, two-third of the active users are involved in social situations. We conclude from the aforementioned statistics a high potential of at least $10\%$ for improving the prediction accuracy of the mobility model based on social influences.

The spatial-temporal (ST) PPM model is able to predict the next location of a user with an accuracy of $18.6\%$. The accuracy using SOST PPM model increases to $21.2\%$ and $22.5\%$ when estimating synchronous specific social influences according to \autoref{eq_pmmSoc} and \autoref{eq_pmmSoc2} respectively. Estimation according to \autoref{eq_pmmSoc2} gives social influences more importance than \autoref{eq_pmmSoc}, hence the better performance. The absolute improvements in accuracy corresponds to 0.026 and 0.039, the relative improvements in accuracy corresponds to 14\% and 21\% respectively.

The drift functions increase the prediction accuracy of SOST PPM to $23.1\% (\beta = 0.02)$ and $23.8\% (\beta = 0.05)$ using both estimators according to \autoref{eq_pmmSoc2} and \autoref{eq_pmmSoc} respectively. The improvements in accuracy correspond to $0.006$ and $0.026$ absolute improvement, and to $2.7\%$ and $12\%$ relative improvement in accuracy respectively. The significance of the improvement in accuracy is confirmed by two-sided unpaired t-tests with ($P(\epsilon) = 2.5*10^{-31}$ and $P(\epsilon) = 0.02$ respectively. The values of $\beta$ imply that social influences decay in three to six weeks.

\autoref{evaluationResult} contains the cumulative improvements in accuracy when incorporating an additional class of social influence factors. The accuracy improves to $19.78\%$, $20.95\%$ and $22.04\%$ by additionally incorporating class I, II and III synchronous specific social influence factors. The total absolute improvement in accuracy by incorporating all classes of synchronous specific social influence factors is $0.0344$, the relative improvement in accuracy corresponds to 18.5\%. The significance of the improvements is confirmed by corresponding two-sided unpaired t-tests (\autoref{evaluationResult}). The empirical impressively underline the importance of social influence factors for enhancing the accuracy of next location prediction.

\begin{table}[htb]
\centering
\begin{tabular}[c]{|l|l|l|l|l|}
\hline
\scriptsize \emph{$\mathfrak{s}$-class} & \scriptsize{\begin{tabular}[l]{@{}l@{}} \emph{Absolute} \\ \emph{Impr. \%.} \end{tabular}} & \scriptsize{\begin{tabular}[l]{@{}l@{}} \emph{Relative} \\ \emph{Impr. \%} \end{tabular}} & \scriptsize{\begin{tabular}[l]{@{}l@{}} \emph{Two-Sided} \\ \emph{Unpaired} \\ \emph{T-Test $P(\epsilon)$} \end{tabular}} \\
\hline \hline
  \scriptsize \emph{Class I:} & \scriptsize 0.0088 (0.0118) & \scriptsize 4.7 (6.1) & \scriptsize{ 0.0012 (0.00113)} \\
  \hline
  \scriptsize \emph{Class I \& II:} & \scriptsize 0.0235 (0.0313) & \scriptsize 12.6 (16.3) & \scriptsize{\begin{tabular}[l]{@{}l@{}} 3.8 * $10^{-28}$ \\ (1.4 * $10^{-43}$) \end{tabular}} \\
  \hline
  \scriptsize \emph{Class I-III} & \scriptsize 0.0344 (0.0458) & \scriptsize 18.5 (23.9) & \scriptsize{\begin{tabular}[l]{@{}l@{}} 1.4 * $10^{-39}$ \\ (1.5 * $10^{-76}$) \end{tabular}} \\
  \hline
\end{tabular}
\setlength{\abovecaptionskip}{1ex}
  \caption{\scriptsize Empirical results: Column 2 represents the absolute improvement in accuracy compared to ST PPM VOMM model, column 3 represents the relative improvement in accuracy compared to ST PPM VOMM model, column 4 the results of two-sided unpaired t-tests (probability of error $p(\epsilon)$) for showing the significance of the improvements. The numbers in braces represent the corresponding values for the portion of users who are involved in at least one social situation (setting $\beta = 0.05$ and $\Delta t = 1$, $\kappa=3$).}

  \label{evaluationResult}
\end{table}

Almost one third of the users were not involved in any social situation, the incorporation of general (trends) social influences is the only possibility of enhancing their next location prediction. The improvement in accuracy increases to $23.8\%$ when additionally general social influences are incorporated. The prediction accuracy increases by almost $0.0178$ due to the incorporation of social influences. The total absolute improvement in accuracy is $0.0522$, which corresponds to a relative improvement in accuracy of $28\%$. The significance of the improvement is confirmed by a two-sided unpaired Student's t-test ($P(\epsilon) = 2.0 * 10^{-19}$). The importance of general social influences is emphasized by considering only users who are not involved in any social situation. The incorporation of general social influences leads to an absolute improvement in accuracy of $0.0221$, which corresponds to a relative improvement in accuracy of $11\%$.

Fano's inequality \cite{fano1961} shows for the users in the Foursquare data-set an average predictability of $29\%$. SOST PPM is able to achieve a prediction accuracy of $23.8\%$ from a maximum predictability of $29\%$, which corresponds to an accuracy of at least $23.8/29 > 82\%$, which impressively underlines the prediction power of SOST PPM. The impact of social influences on improving location prediction can be convincingly demonstrated considering only users who were involved in only one social situation (The numbers in braces in \autoref{evaluationResult}), the improvement in accuracy increases to approximately $\approx 0.0615$ which corresponds to a relative improvement in accuracy of $32\%$.

The users in the data-set visit in 437\,231 cases (unknown) locations which are not yet been seen in their own location history. Integration of social networks resulted to a correct prediction of the location of users in 36 \, 469 of these cases. The absolute improvement in total prediction accuracy is 0.0319, which corresponds to $61\%$ of the total absolute improvement in accuracy. The significance of the improvement is confirmed by a two-sided unpaired Student's t-test ($P(\epsilon) = 8 * 10^{-11}$). 

\begin{figure}[htb]
\centering
 \includegraphics[width=0.5\textwidth]{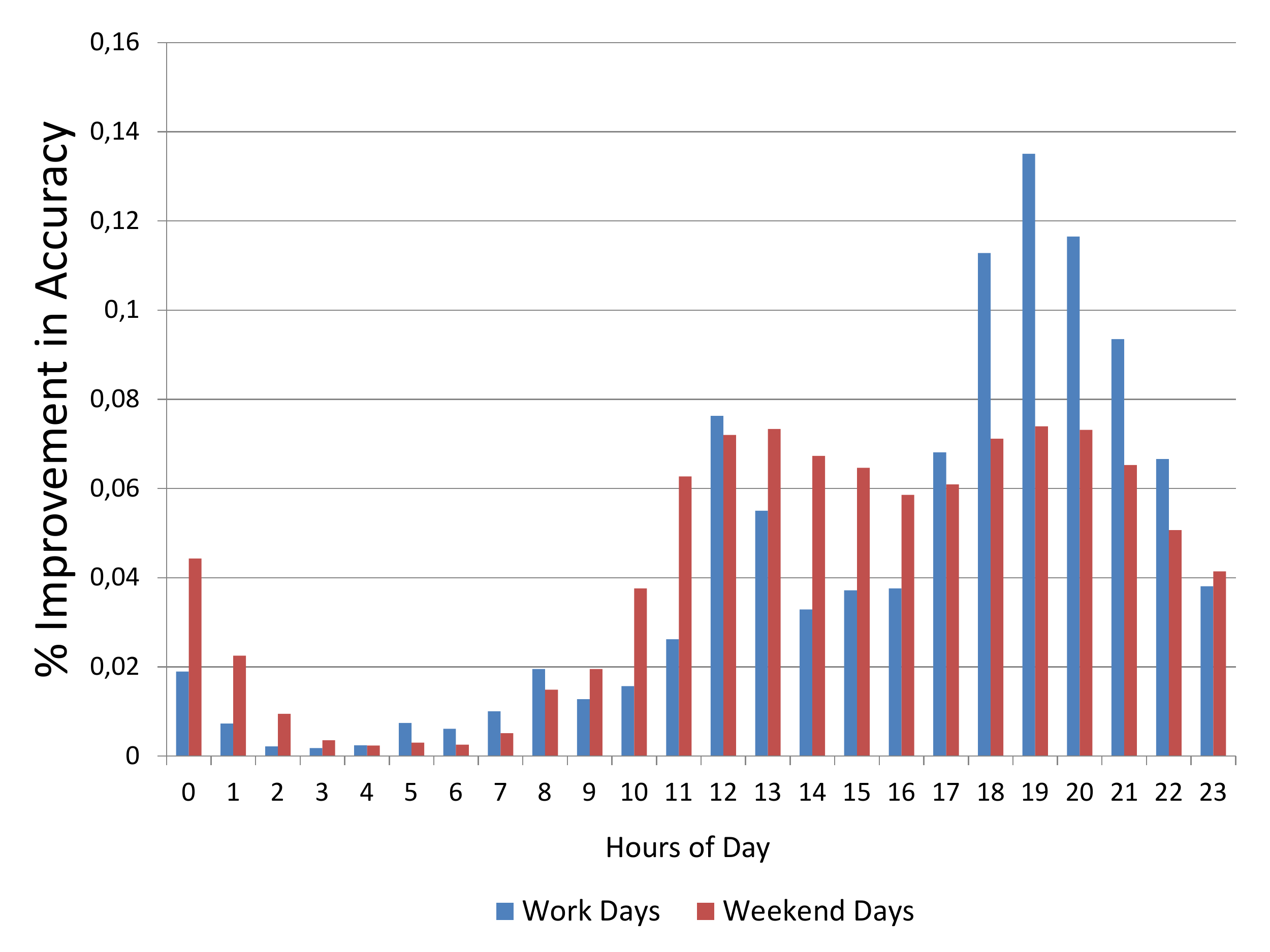}

\setlength{\abovecaptionskip}{0ex}
\caption{\label{ImprovementsHoursOfDay}
\footnotesize The proportion of improvement in accuracy over the hours of weekend days (red bars) and work days (blue bars).}
\end{figure}

The uncertainty in predicting the next location of a mobile user is the highest during evening hours of working days and on weekend days when they spend time with recreational activities. \autoref{ImprovementsHoursOfDay} shows that the most improvement in accuracy due to social influences occurs during these time periods. The most improvement in accuracy on work days (blue bars) is achieved lunch and evening hours. has two peaks. The most improvements on weekend days (red bars) covers the hours between 11 a.m. and 12 p.m. These are the typical periods where people spend time with their friends, for example for having lunch, a drink after work, a window shopping stroll around the city. \autoref{ImprovementsHoursOfDay} again underlines the importance of social influences for improving the prediction accuracy during time periods with high uncertainty, when the individual mobility model fails to find mobility patterns.

\subsubsection{Social Network Measurements}

\begin{figure}[htb]
\centering
 \includegraphics[width=0.5\textwidth]{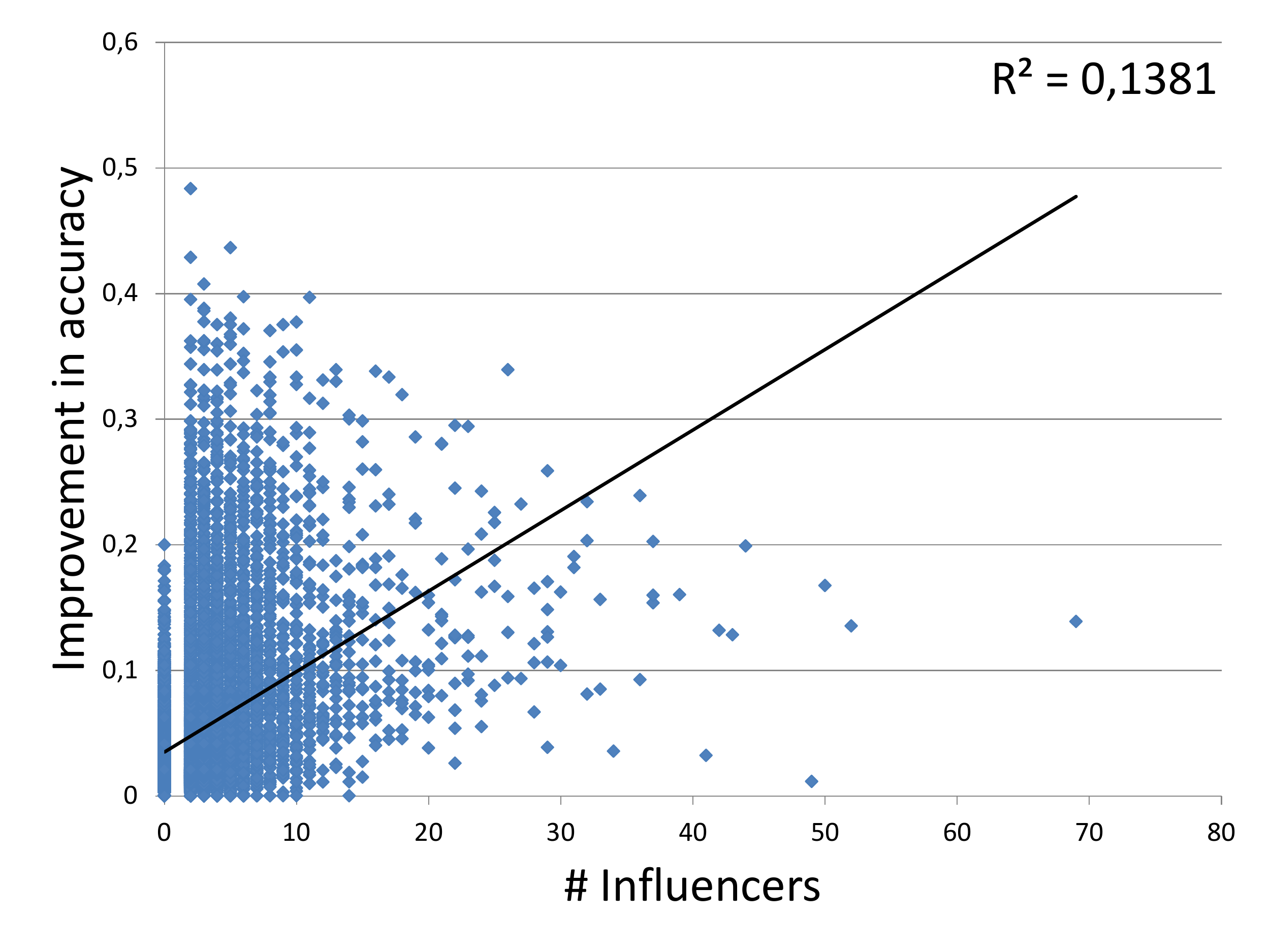}

\setlength{\abovecaptionskip}{0ex}
\caption{\label{improvementInfluencer}
\footnotesize The relationship between the number of influencers (y-axis) and absolute improvement in accuracy (x-axis) (Setting $\Delta t$ to one hour).}
\end{figure}

The mobility of most of the users (followers) in the data-set is influence by a small subset of their friends (influencers). Fewer than $1\%$ of the users have more than 20 potential influencers. The inclusion of location histories of only two friends is sufficient to achieve a significant relative improvement in accuracy of $21\%$. The relationship between the number of influencers and the improvement in accuracy shows a strong positive trend, which is confirmed by a moderate positive correlation coefficient according to Pearson's correlation coefficient ($r=0.37$) and a strong positive correlation according to Spearman's rank correlation coefficient ($\rho=49$) with a probability of error of zero ($\epsilon=0.0$) (\autoref{improvementInfluencer}).

The size of location histories injected by the influencers exhibits a similar trend as the number of influencers. The inclusion of only 50 visits by influencers is sufficient to improve the accuracy by a significant $0.0266$ absolute improvement and $14\%$ relative improvement. The improvement in accuracy shows a positive trend with the size of injected location histories of friends. The positive trend is confirmed by a moderate positive correlation of $0.23$ according to Pearson's correlation coefficient, and a similar positive correlation of $0.21$ according to Spearman's correlation coefficient with a probability of error of zero ($\epsilon=0.0$).

\begin{figure}[htb]
\centering
 \includegraphics[width=0.5\textwidth]{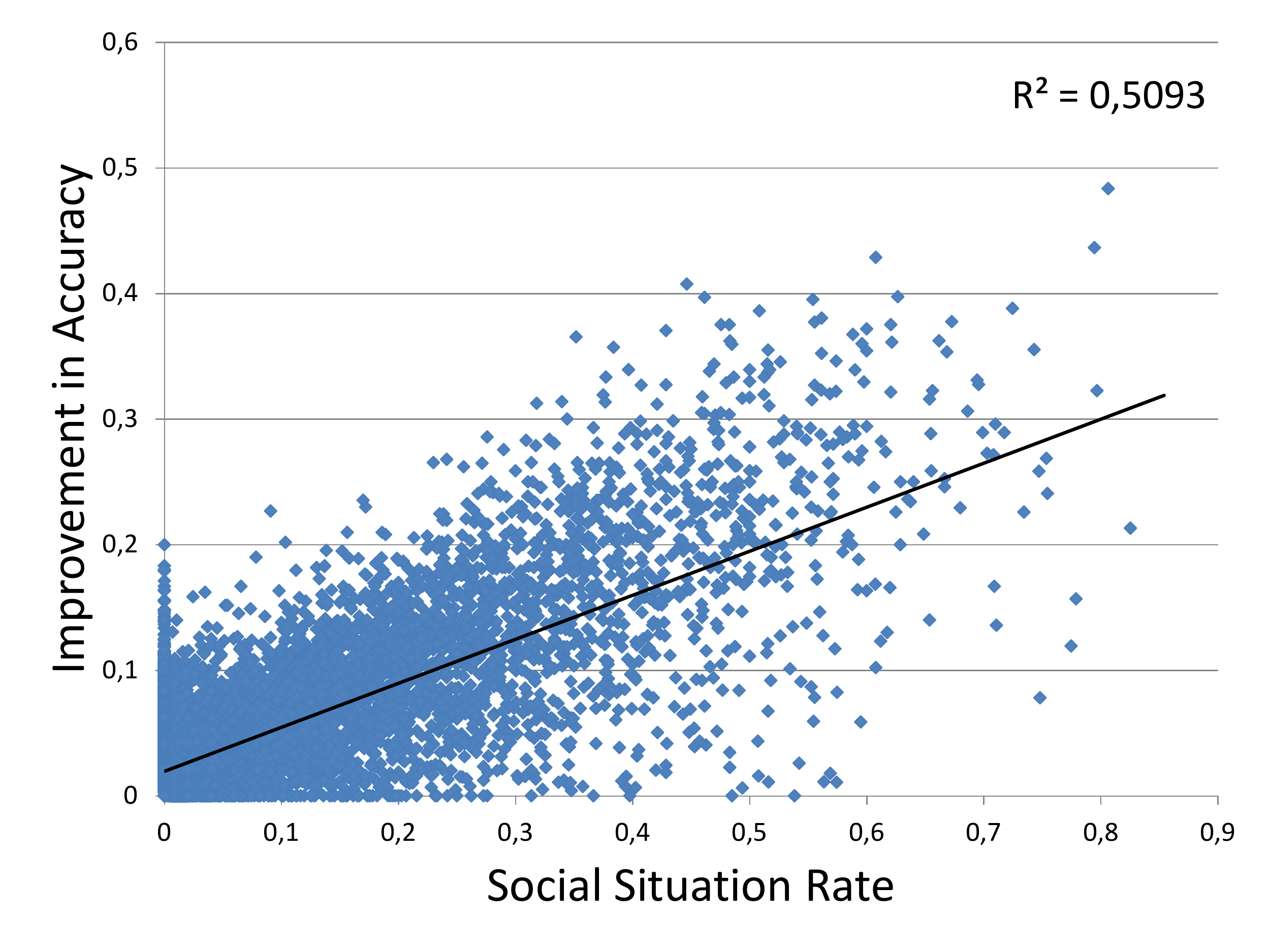}

\setlength{\abovecaptionskip}{0ex}
\caption{\label{improvementSocSitRate}
\footnotesize Absolute accuracy improvement correlates with the average social situation rate $r = 0.71, \rho = 0.61, P(\epsilon) = 0.0$.}
\end{figure}

Outgoing, talkative, energetic behavior manifests the character of extraverted users, whereas reserved and solitary behavior manifest the character of introverted users \cite{Thompson2008}. Expectedly, extrovert users are involved in more social situations compared to introvert users, hence, their mobility behavior is more predictable via social amendments to the spatial-temporal ST PPM approach. The relationship between social situation rate and improvement in accuracy on \autoref{improvementSocSitRate} shows a positive trend that confirms this behavior. The positive trend is underlined by a very strong correlation according to both Pearson's $0.71$ and Spearman's $0.61$ correlation coefficients.

The number of social situations that have been integrated in the SOST tree is $149\,700$. Almost $70\%$ of these social situations are among members of the same 2-plexes, thus most social influences is transferred between members of the same cohesive subgroup, which is in accordance with the results of the correlation analysis in section \ref{corrCohesion}.

\begin{figure}[htb]
\centering
 \includegraphics[width=0.5\textwidth]{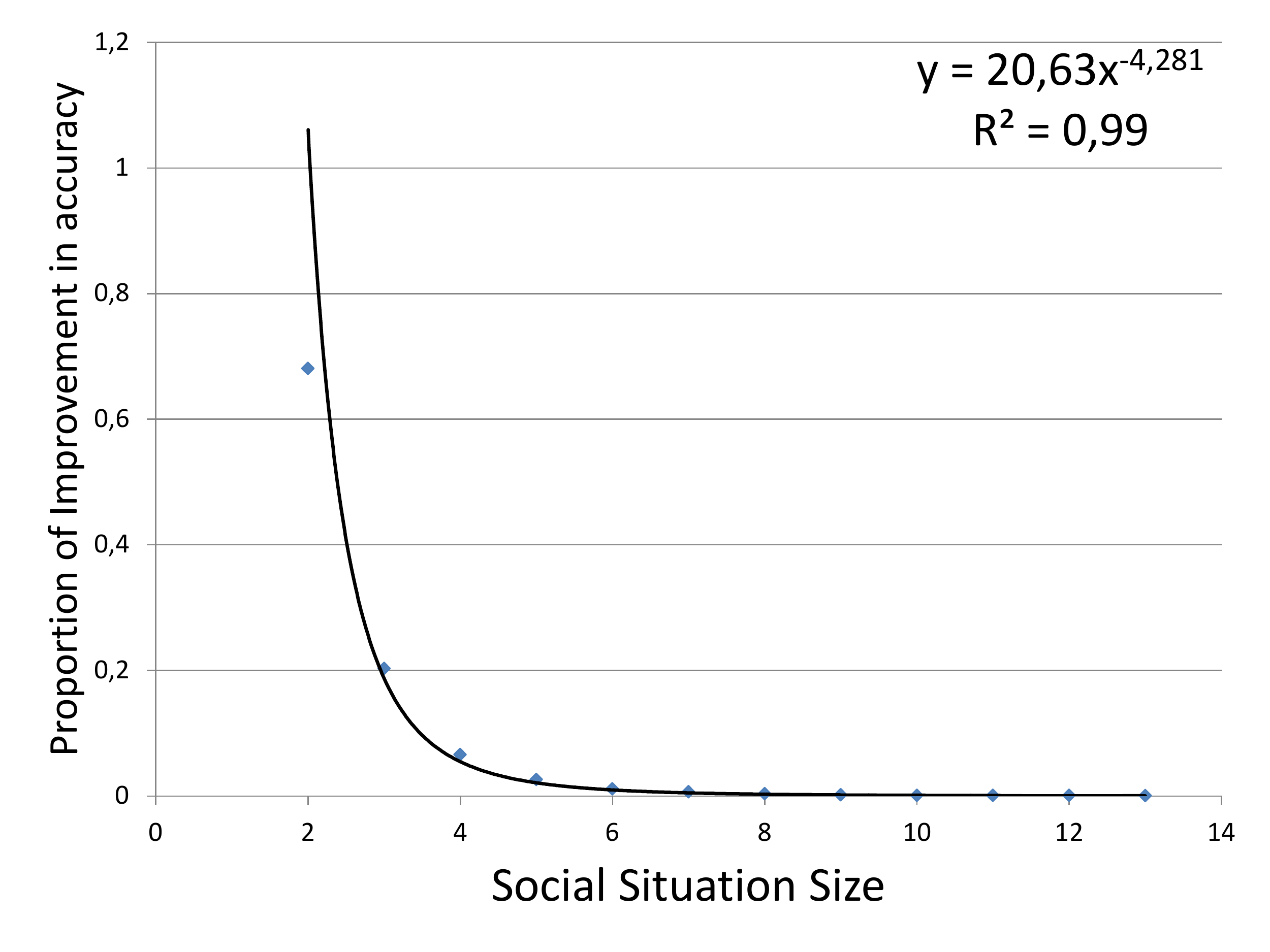}

\setlength{\abovecaptionskip}{0ex}
\caption{\label{improvementSSSize}
\footnotesize The relationship between the percentage of total absolute improvement in accuracy and the size of social situations follows a power law with a coefficient of determination of $0.99$.}
\end{figure}

\begin{figure}[htb]
\centering
 \includegraphics[width=0.5\textwidth]{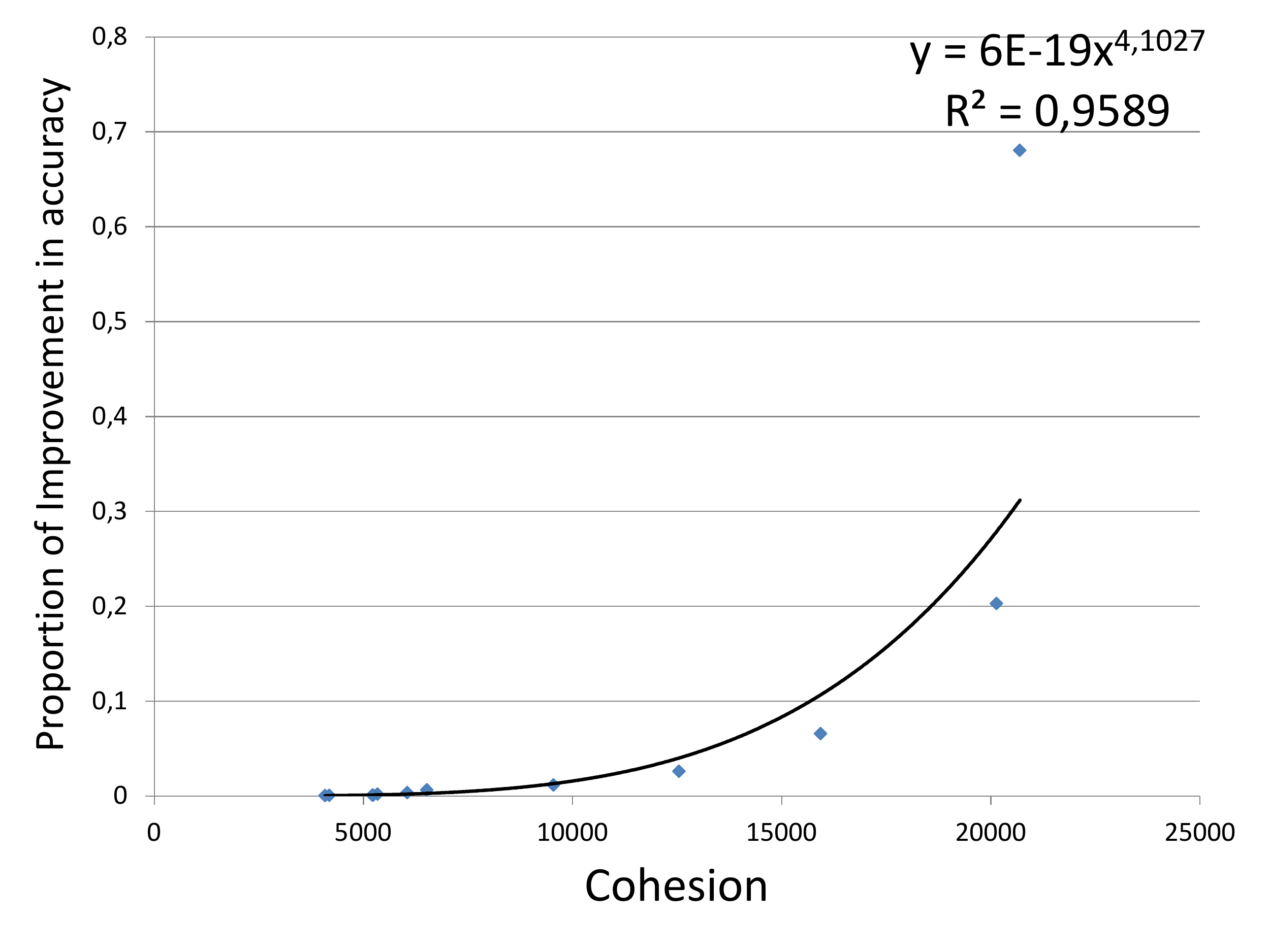}

\setlength{\abovecaptionskip}{0ex}
\caption{\label{improvementCohesion}
\footnotesize The relationship between the percentage of total absolute improvement in accuracy and the average measure of cohesion in the social situations follows a power law with a coefficient of determination of $\backsimeq 0.96$.}
\end{figure}

The size of a social situation is defined by the number of involved users. The size of most social situations varies between two and five, and only a small portion are larger. The relationship between the size of social situations and the cohesion according to \autoref{measureOfCohesion} follows power law, the higher the size of social situations, the lower the cohesion. The relationship between improvement in accuracy and both size and measure of cohesion of social situations follow pow lows with coefficients of determination of $R^2 > 99$ (\autoref{improvementSSSize}) and $R^2 > 95$ (\autoref{improvementCohesion}) respectively. Improvement in accuracy shows a strong positive correlation with the measure of cohesion in accordance with the correlation analysis in section \ref{corrCohesion}.

Humans have spatial, temporal, cognitive, emotional limitations, that prevents them from maintaining all their relationships with the same intensity \cite{Granovetter2005}. Dunbar suggests the number of neighbors, with whom a user can maintain stable cognitive social relationships to be 150 (\cite{Dunbar1992} as cited by \cite{DunbarsNumber}. Almost $9\%$ of the users in the Foursquare dataset have more than 150 neighbors, and 22 of the users have at least 1\,000 neighbors, which means that these users have a lot of weak ties, because intuitively no one can cognitively maintain such a number of relationships. A user in a social network maintains strong ties to a small subset (2-plex) of their friends, most of whom are in touch with one another \cite{Granovetter2005}, and weak ties to the remaining friends, let us say acquaintances (in accordance to Granovetter). The acquaintances in turn have their own subsets of strong ties, thus weak tie bridges the gap between different communities and social circles \cite{Granovetter1982} and are important for transmitting general social trends beyond the borders of a cohesive subgroup. The information of users in a cohesive subgroup overlap to high degree due to the intensity of their interaction, which results in  homogeneity in their behavior, life styles, emotional needs, thoughts, beliefs, movements, goals, etc. Two users connected via a weak tie exchange rather more novel information \cite{Granovetter2005} because of the heterogeneity in their information. The heterogeneity occurs because each of the two users spends time and interacts with people, who the other user does not know \cite{Granovetter2005}.

Degree centrality is a notion that refers to the extent in which a user is connected to others. Gladwell refers to Central users with connectors, a few people who have the extraordinary knack of making friends and acquaintances and who can bring users from different social circles together (\cite{Gladwell2002} ~Pages 38-41 as cited by \cite{tippingPoint}). Central users have more friends than they cognitively can maintain strong relationship. Hence, most of their ties are rather weak ties that are enmeshed in different cohesive subgroups. Therefore, Central users are important, because they can bridge the gap between many different social communities and thus transmit social influence between these communities.

\begin{figure}[htb]
\centering
 \includegraphics[width=0.5\textwidth]{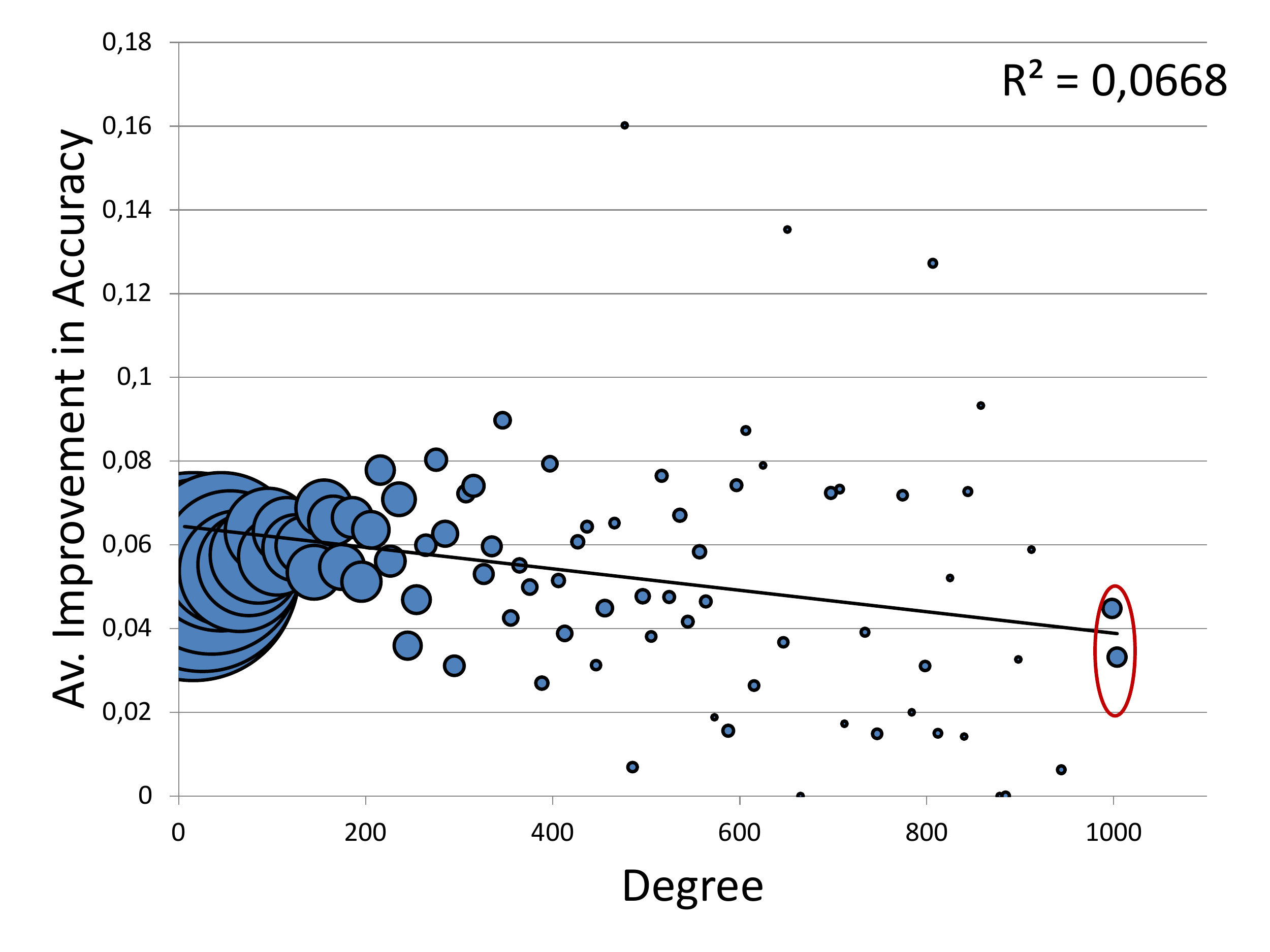}

\setlength{\abovecaptionskip}{0ex}
\caption{\label{improvementDegree}
\footnotesize The average absolute improvement in accuracy shows a negative trend as the degree increases, the correlation coefficients were found to be ($r = -0.26, \rho = -0.29, P(\epsilon) = 0.0$).}
\end{figure}

The Foursquare dataset contains 147,900 social situations, almost $> 70\%$ of the social situations are between members of the same maximal 2-plexes, in the remaining $30\%$ cases the users interact with their acquaintances. We enclosed the (central) users with more than 1\,000 (many weak ties) in red ellipses in the following three figures. \autoref{improvementDegree} represents the relationship between the average degree and the improvement in accuracy. The figure shows that the prediction accuracy of central users is only slightly improved using the location histories of their friends, because a central user interacts more with their weak ties from different social communities. Nevertheless, central users are important for transmitting social influence, because a check-in of a central user can potentially influence the mobility of 1,000 neighbors. Central users are trend setters or trend transmitters between different social communities and are followed by rather than following others.

\begin{figure}[htb]
\centering
 \includegraphics[width=0.5\textwidth]{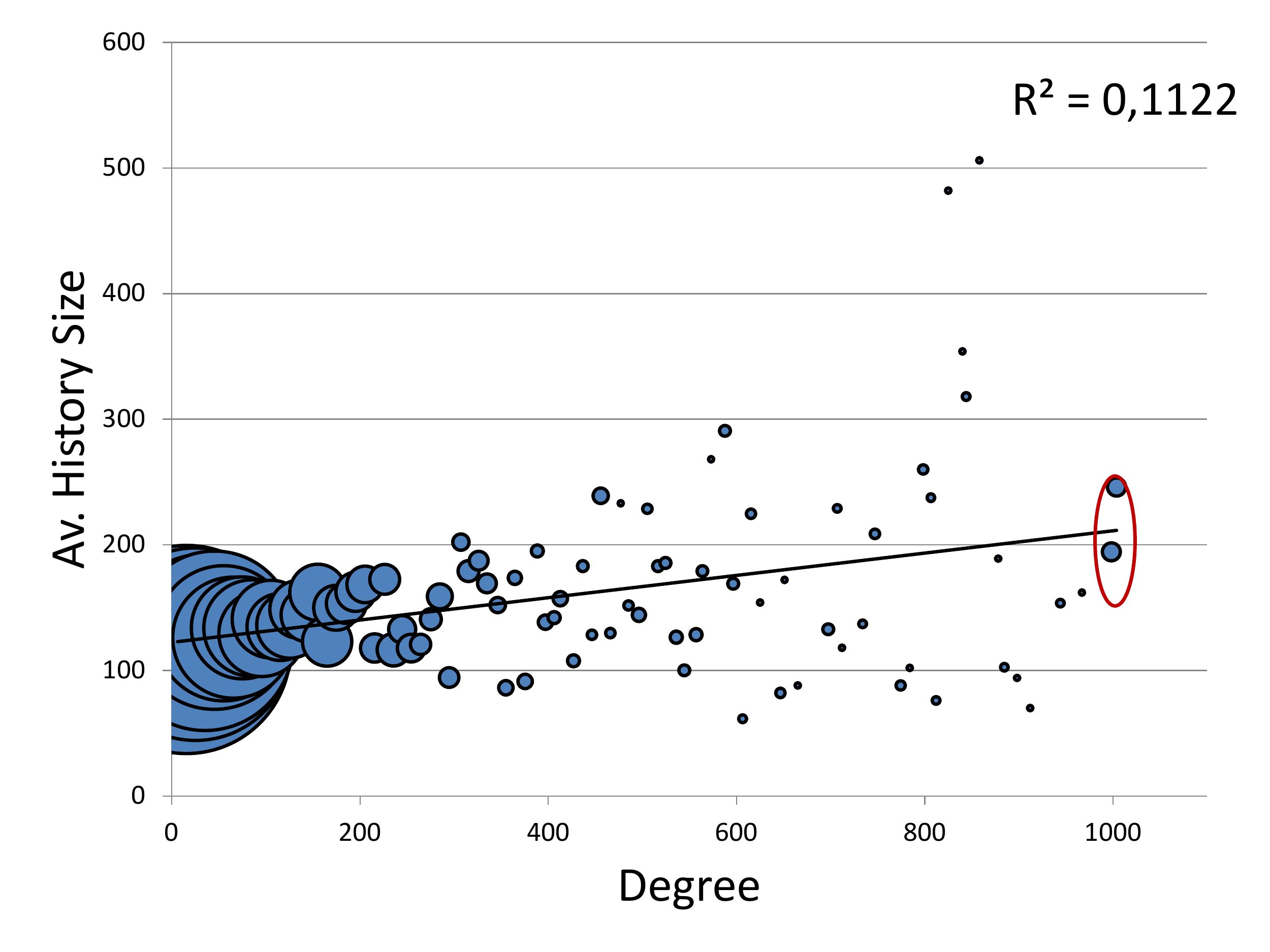}

\setlength{\abovecaptionskip}{0ex}
\caption{\label{degreeHistory}
\footnotesize A plot showing the positive trend between the degree and the average location history ($r = 0.33, \rho = 0.25, P(\epsilon) = 0.02$). The size of the bubbles indicates to the number of users with a given degree.}
\end{figure}

\begin{figure}[htb]
\centering
 \includegraphics[width=0.5\textwidth]{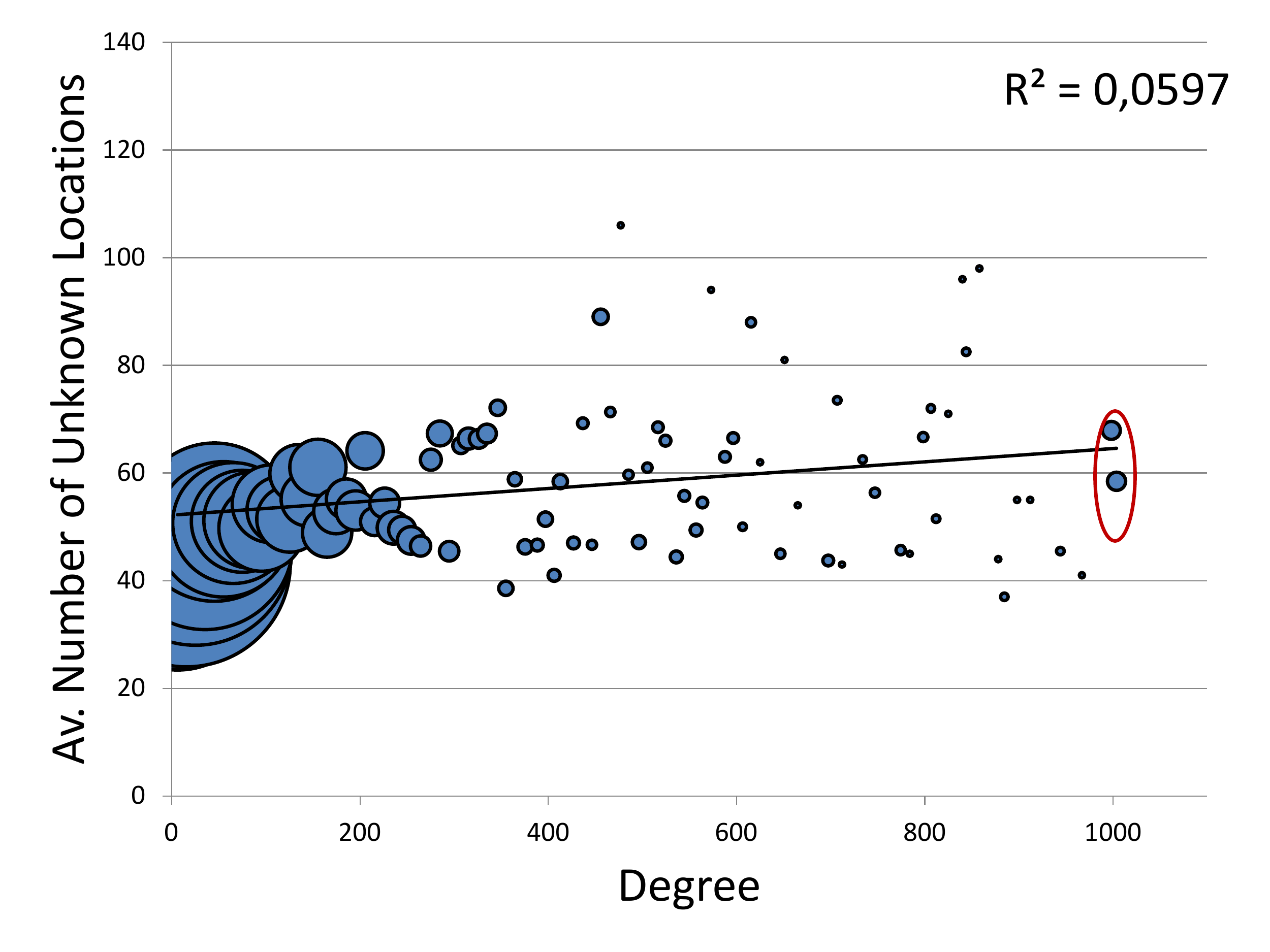}

\setlength{\abovecaptionskip}{0ex}
\caption{\label{degreeUnknownLocations}
\footnotesize A plot showing the positive correlation between the degree and the average number of locations visited for the first time ($r = 0.24, \rho = 0.21, P(\epsilon) = 0.05$). The size of the bubbles indicates to the number of users with a given degree.}
\end{figure}

\ref{degreeHistory} shows the relationship between degree and average size of locations. The plot shows a positive trend $r = 0.16, \rho = 0.15, P(\epsilon) = 0.00016$. The positive trend states that central users get about a lot, and are explorative in nature. They are trend setters and can influence the mobility of their neighbors. \autoref{degreeUnknownLocations} shows a positive trend between degree and average number of locations visited for the first time $r = 0.33, \rho = 0.25, P(\epsilon) = 0.0$ confirming the explorative nature of central users. The above results underline the importance of central users in a social network for transferring influence and their contribution to improve the prediction accuracy of their neighbors.

\subsubsection{Location History Measurements}

The performance of predictive models based on probabilistic reasoning depends to a high extent on the existence of sufficient location history. The inclusion of locations histories of friends helps adding a vast amount of information not observed by the user. It seems to be obvious that the inclusion of social networks improves prediction accuracy rather for users with insufficient (small) location histories. But, the relationship between improvement in accuracy and history size shows no tendency, because both entropy and number locations visited by the user increase as the size of location history increases. Therefore, social influences will always remain indispensable for enhancing prediction accuracy, regardless the size of location history of the user in question for location prediction.

\begin{figure}[htb]
\centering
 \includegraphics[width=0.5\textwidth]{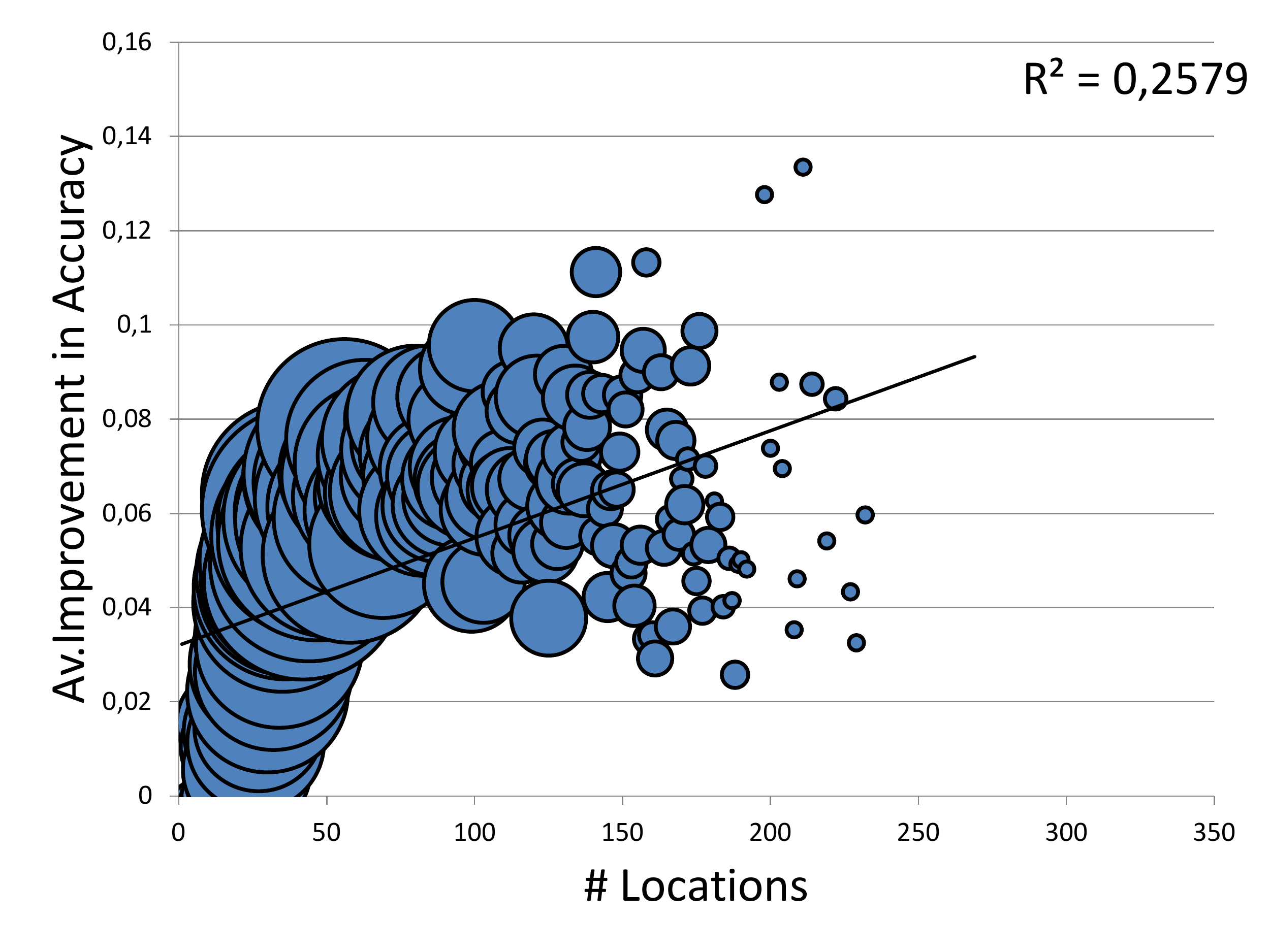}

\setlength{\abovecaptionskip}{0ex}
\caption{\label{improvementLocationAverage}
\footnotesize The plot shows a positive correlation between the number of locations visited by each user and average absolute improvement in accuracy $r = 0.51, \rho = 0.42, P(\epsilon) = 0.0$.}
\end{figure}

Users who are explorative in nature visit a lot of locations with similar probabilities, thus their mobility is less predictable. The users visits in $38\%$ of the cases new locations, but most of these locations are previously visited by their friends. Therefore, injecting location histories of friends into the individual mobility of a user can indeed increase the prediction accuracy. \autoref{improvementLocationAverage} shows a positive trend between average number of locations visited by the different users and the average improvement in accuracy. The positive trend is confirmed by a strong positive correlation coefficient according to Pearson $r = 0.51$ and Spearman's $\rho = 0.42, p(\epsilon) = 0$.

\begin{figure}[htb]
\centering
 \includegraphics[width=0.5\textwidth]{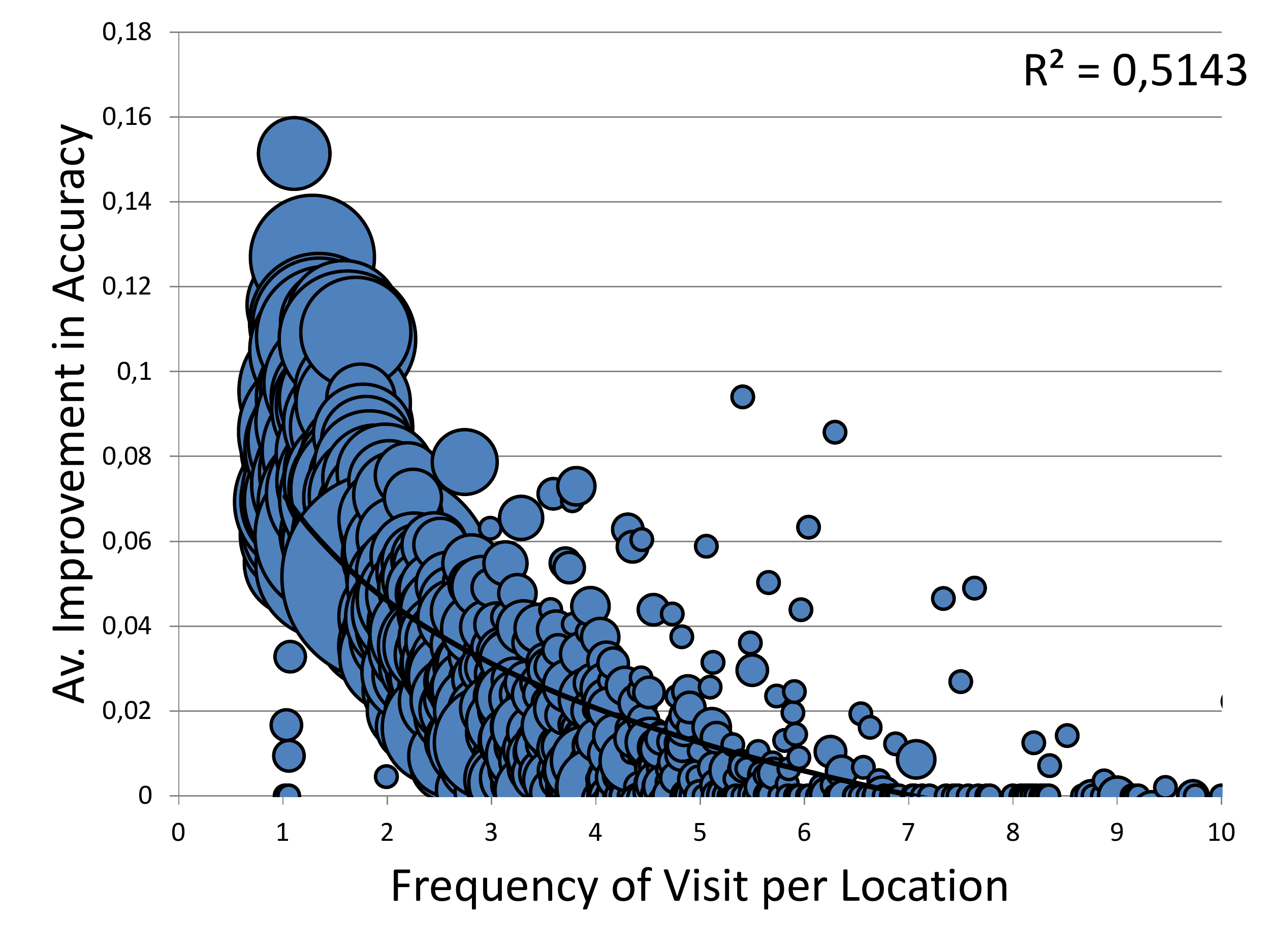}

\setlength{\abovecaptionskip}{0ex}
\caption{\label{improvementHistoryLocationAverage}
\footnotesize The plot shows a negative correlation between the frequency of visit per location and average absolute improvement in accuracy $r = -0.34, \rho = -0.80, P(\epsilon) = 0.0$.}
\end{figure}

Average frequency of visit per location is a measure that affects the prediction accuracy. The mobility of users with low average frequency of visit per location is less predictable. The incorporation of location histories of friends into the individual mobility model of a user helps increase prediction accuracy. The average frequency of visit per location in the Foursquare dataset is very low, each user makes on average $2.04$ check-ins per location. \autoref{improvementHistoryLocationAverage} underlines the importance of social networks for increasing prediction accuracy, the relationship between average frequency of visit per location and average improvement in accuracy due to the integration of location histories of friends shows a strong negative trend. The negative trend is confirmed by a strong negative correlation coefficient of $r = -0.34$ according to Pearson, and a very strong negative correlation coefficient of $\rho = -0.80, P(\epsilon)=0$ according to Spearman.

\begin{figure}[htb]
\centering
 \includegraphics[width=0.5\textwidth]{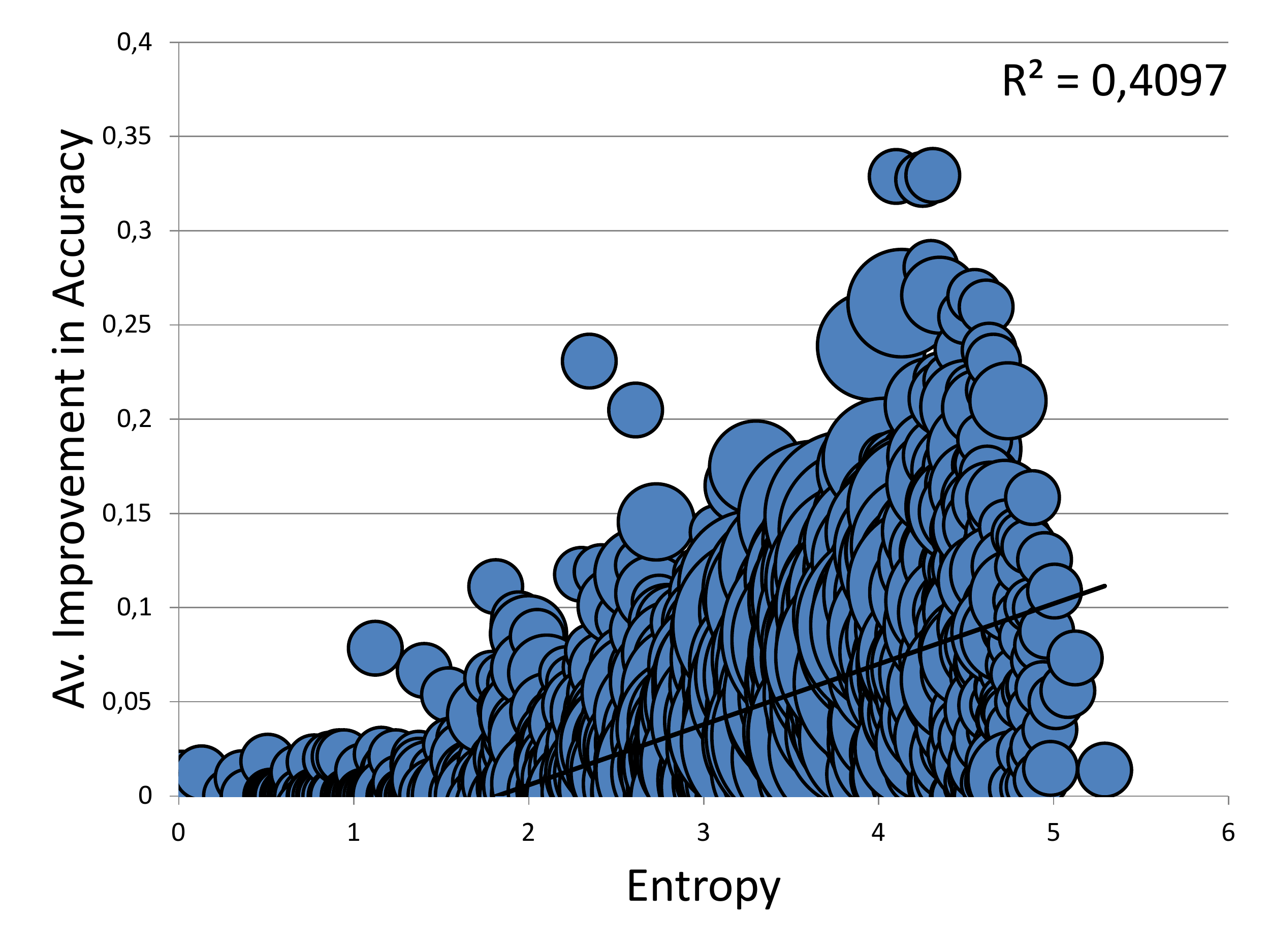}

\setlength{\abovecaptionskip}{0ex}
\caption{\label{improvementEntropyAverage}
\footnotesize Average absolute improvement in accuracy shows a positive trend with increasing entropy of the users. Both Pearson's and Spearman's correlation coefficients are found to be $r = 0.64, \rho = 0.72, P(\epsilon) = 0.0$ respectively.}
\end{figure}

The importance of social networks for location prediction is again underlined by investigating the relationship between entropy and improvement in accuracy due to the integration of location histories of friends into the individual mobility of a user. Entropy is a measure for the uncertainty associated with predicting the next location of a mobile user. The high average entropy value of $3.48$ is an indicator for low mobility predictability in the Foursquare data-set. \autoref{improvementEntropyAverage} shows a strong positive trend between entropy and improvement in accuracy due the integration of social networks. The strong positive trend is confirmed by very strong correlation coefficient values of $r = 0.64$ and $\rho = 0.72, P(\epsilon) = 0.0$ according to Pearson and Spearman respectively.

\begin{figure}[htb]
\centering
 \includegraphics[width=0.5\textwidth]{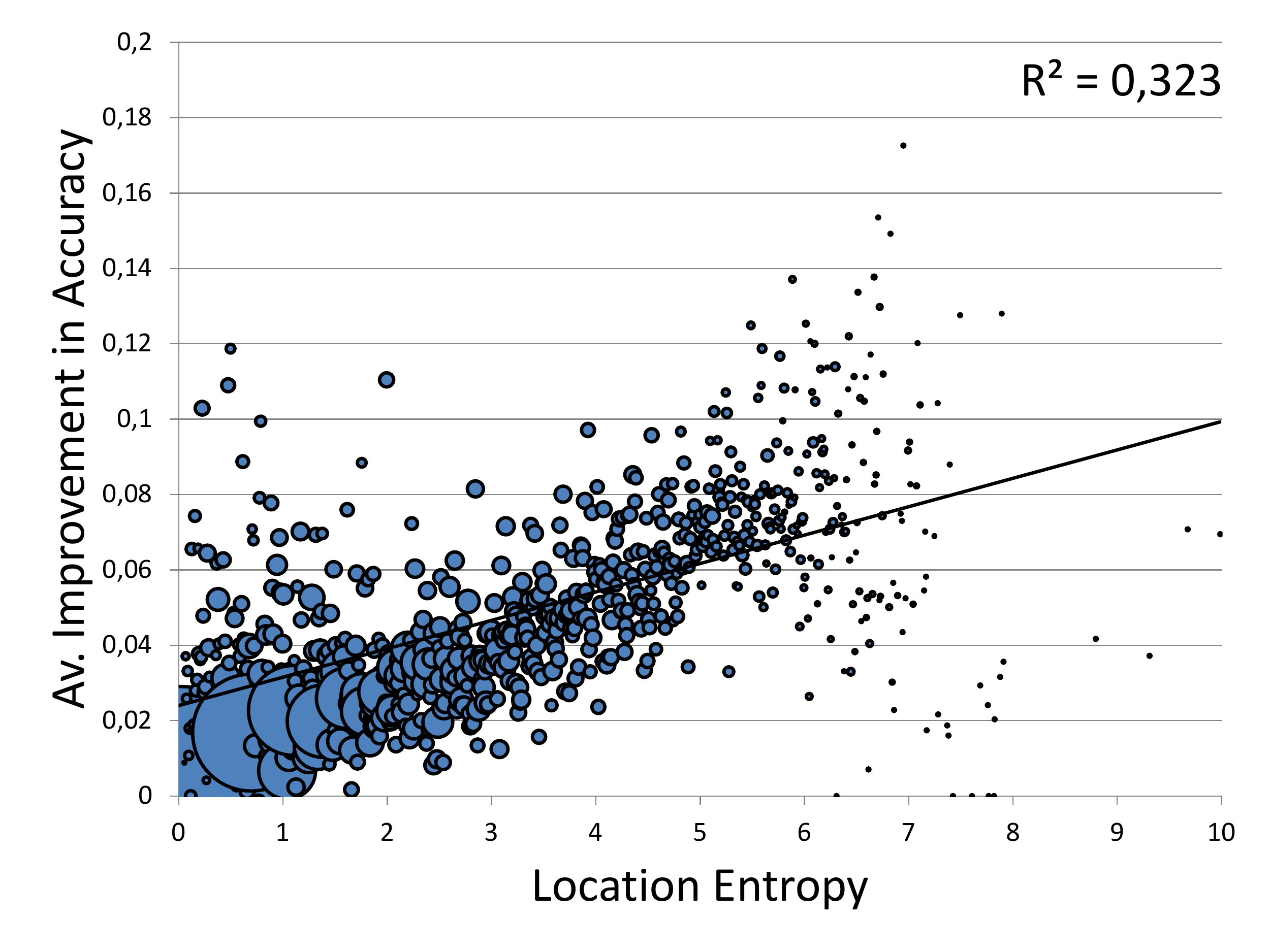}

\setlength{\abovecaptionskip}{0ex}
\caption{\label{improvementLocationEntropyAverage}
\footnotesize Average absolute improvement in accuracy shows a positive tendency with the increasing entropy of the users. Pearson's and Spearman's correlation coefficients were found to be  $r = 0.57, \rho = 0.60, P(\epsilon) = 0.0$ respectively.}
\end{figure}

Location entropy is a measure of predictability of locations. Locations visited by many users with similar frequencies are highly entropic and their predictability is associated with high uncertainty. Examples of highly entropic locations are airports, sport stadiums, underground stations, etc. A restaurant visited frequently by neighboring residents and sporadically by visitors from elsewhere is an example of mediocre entropic locations. A private domicile of a user where friends come by occasionally is an example of low entropic locations. \autoref{improvementLocationEntropyAverage} shows a strong positive trend between location entropy and improvement in accuracy due to the integration of social networks. The strong positive trend is confirmed by very strong correlation coefficient values of $r = 0.57$ according to Pearson, and $\rho = 0.60, P(\epsilon) = 0.0$ according to Spearman. Integration of social network helps reduce the uncertainty associated with predicting high entropic locations.

\subsubsection{Mobility Models Based on Discrete HMMs}

The Hidden Markov Model (HMM) is probably one of most popular, well-studied and powerful DBN approach (\ref{sectionHMM}). Learning the exact parameters of an HMM and the complexity of the solution represents an intractable task, because as none of the known learning algorithms can find an exact solution, they tend to find a solution with local maxima \cite{Ron96thepower}. Therefore, selection of the initial parameters of an HMM is of immense importance. The selection of proper initial model parameters help reduce convergence time and to increase the probability of convergence to a true solution. Therefore, HMM modeling requires in-depth understanding of the application domain.

In \cite{arthurGrohBapierre}, two different two states HMMs have been constructed, one model with and the other without considering influences from social networks. The accuracy of both models ($16.4$ \& $16.6$) is considerably lower than the corresponding PPM VOMM mobility models ($18.6$ \& $23.8$) in the previous section. The poor performance of HMM is due to the difficulty of learning the model parameters. The performance of HMM decreases as the the number of states in the model increases (for example to $15$ \& $15.2$ when using five states), probably because of one of two reasons, either the data has been drawn from one distribution, or the location histories of the users are insufficient to learn the exact model parameters. We assume the second reason is rather probable, because observation size is known to be a critical issue for HMM \cite{Ron96thepower, begleiter2004}. The availability of more location history results not necessarily increasing the performance of model, because as we have stated earlier, both entropy and the number of locations visited by a user increases as the history size increases. The increasing entropy and number of locations may lead either to a changed number of states in the model and thus make re-training the model indispensable (which is another critical issue of HMM) or it leads to severe readjustment of the model parameters, especially when frequently occurring states are no longer relevant for a user (for example, behavioral changes due to marriage, moving or a new job, etc.).	

\section{Conclusion}
%


The rapid technological advances of the last years, especially the pervasiveness of mobile devices such as smart-phones, as well as the spread of mobile access to the Internet and the emergence of social networking platforms allow the collection of vast amounts of data containing information about the behavior of users, that facilitates the investigation of human mobility. In this work, using a data-set from an online location based social networking platform "Foursquare", we investigated the existence of statistical interdependence between human mobility and social proximity, as well as the impact of social networks on influencing the mobility behavior of mobile users.

The empirical results show indeed a strong interdependence between social proximity and mobile homophily. An in depth correlation analysis between different social proximity measures such as common neighbors, Jaccard coefficient, Adamic \& Adair on one hand side, and different mobile proximity measurements such as co-location count, social situation rate, spatial cosine similarity etc. on the other hand, has confirmed this interdependence. Further, using a influence model based on variable markov model, we have shown that impacts from the social network indeed cause changes in the mobility behavior of individual. We investigated the causation effect by improving location prediction of an individual by incorporating the location histories of their friends. The absolute improvement in accuracy was 5.2\%, the relative improvement even 28\%.

Privacy concerns are of great relevance regarding the acceptance of users towards location based (social network) services LB(SN)S. A service gains a higher acceptance if users have the choice to opt-in or opt-out of a service, if they know who has access to their location information and with whom they share their information and for how long \cite{Perusco2007}. The empirical results has shown that the influence of movements of friends approaches zero after three to six weeks. Further, the inclusion of locations histories of few friends is sufficient for enhancing location prediction significantly. This finding is important for increasing user acceptance towards location prediction when they know, that they need to share their location histories with a small subset of their close friends for a comparably limited duration less than six weeks.

The empirical results has shown that the mobility of an individual is influences mostly by the members of the same cohesive subgroups and that cohesion inside the groups shows a very strong correlation with improvement in accuracy. The members of the same cohesive subgroup are responsible for transferring social influence/trends inside the groups. The members of a cohesive subgroup exhibit similarities in their goals, believes, information, emotional needs, interests etc. resulting in a high similarity in their information and behavior. The users in a social network have more ties than the ties to the members of their cohesive subgroups. Information exchange (co-locations, social situations) between two users connected via a weak tie are responsible for the transmission of more novel information between different cohesive subgroups, because each of the two users are members in different cohesive subgroups, thus their information differ from each other. A central user in a social network (degree centrality) have a lot of weak ties, thus they are very important for the spread of social influence between different social communities and setting new social trends.
 
Location prediction can be used in a variety of services such as optimizing fuel consumption and reduction of co2 emission in vehicles \cite{Ganti2010, Ericsson2006, Lee2008}, increasing driving efficiency \& safety \cite{Wevers2010}, increasing the performance of high-voltage battery pack in hybrid electric vehicles (HEV) (\cite{pesaran2003cooling} as cited by \cite{Krumm2011}), mobile marketing and intelligent mobile advertising \cite{Barnes2004, Barwise2002}, saving energy in private households \cite{Gupta2009}, getting a head for the demand curve and being more proactive in deploying aid and rescue capabilities during disaster relief scenarios (\cite{Gao2011, Gao2011b} as cited by \cite{Gao2012}), rehabilitation or crime suppression using electronic monitoring, healthcare monitoring systems \cite{Minkyong2011},  spread of human and electronic viruses, city planning, resource management in mobile communications\cite{Song2010}, traffic management and public transport recommender systems \cite{Rodriguez2012}, presence prediction for face-to-face meeting, emergency call or intelligent postal services \cite{Krumm2011}, supporting assisting technologies for disabled or cognitively impaired persons (Alzheimer's) \cite{Patterson2002}.

\bibliographystyle{abbrvGeorg}
\bibliography{georgBibFile,halgurtBibFile,arthurBibFile}
\end{document}